\documentclass[a4paper,11pt]{article}

\usepackage{jheppub}
\usepackage{graphicx}
\usepackage{latexsym}
\usepackage{verbatim}

\usepackage{xcolor}

\usepackage{cancel}

\usepackage[normalem]{ulem}

\newcommand{\support}{ \rule[-9pt]{0pt}{24pt} }
\newcommand{\Raise}[2]{ \raisebox{#1}[0pt][0pt]{#2} }
\newcommand{\Dp}{D$p$}
\newcommand{\Dppf}{D$(p+4)$}
\newcommand{\PP}[2]{ \frac{\partial {#1}}{\partial {#2}} }

\newcommand{\DP}{ \Delta_+ }
\newcommand{\DM}{ \Delta_- }
\newcommand{\DS}{ \Delta_* }
\newcommand{\bGG}{ b_{GG} }
\newcommand{\bGC}{ b_{GC} }
\newcommand{\bCG}{ b_{CG} }
\newcommand{\bCC}{ b_{CC} }
\newcommand{\xb}{ \bar{x} }
\newcommand{\bb}{ \bar{b} }
\newcommand{\vphib}{\varphi } 
\newcommand{\Phib}{\Phi } 

\subheader{\hfill USTC-ICTS-15-06}

\title{\boldmath Phase structures of the black \Dp-\Dppf-brane system in various ensembles II:
electrical and thermodynamic stability}

\author[a,b]{Zhiguang Xiao}
\author[a,c]{and Da Zhou}

\affiliation[a]{The Interdisciplinary Center for Theoretical Study,\\
University of Science and Technology of China,\\
Hefei, Anhui, 230026, China}

\affiliation[b]{State Key Laboratory of Theoretical Physics,\\
Institute of Theoretical Physics,\\
Chinese Academy of Sciences, Beijing 100190, China}
  
\affiliation[c]{Department of Mathematics, City University London,\\
London EC1V 0HB, U.K.}

\emailAdd{xiaozg@ustc.edu.cn}
\emailAdd{da.z.zhou@gmail.com}

\abstract{By incorporating  the electrical stability condition into
the discussion, we continue the study on the thermodynamic phase
structures of the D$p$-D$(p+4)$ black brane in  GG, GC, CG, CC
ensembles defined in our previous paper \cite{zhou:2015}. We find that
including the electrical stability conditions in addition to the
thermal stability conditions does not modify the phase structure of
the GG ensemble but puts more constraints on the parameter space where
black branes can stably exist in GC, CG, CC ensembles. In particular,
the van der Waals-like phase structure which was supposed to be
present in these ensembles when only thermal stability condition is
considered would no longer be visible, since the phase of the small
black brane is unstable under electrical fluctuations. However, the
symmetry of the phase structure by interchanging the two kinds of
brane charges and potentials is still preserved, which is argued to be
the result of T-duality. }


\arxivnumber{}

\begin{document}

\maketitle

\section{Introduction}

Investigating the thermodynamic properties and their origins of black holes has been
an over-40-year's industry since Bekenstein found the analogue of the principle of increasing
entropy in the black hole context in 1973 \cite{bekenstein:1973}. Hawking confirmed
black holes do have temperature by applying quantum mechanics to a classical gravitational
background \cite{hawking:1975}. It has been a consensus since then that understanding the
thermodynamic nature of black holes needs a more fundamental theory of quantum gravity
which treats gravitation itself as a dynamical quantum object instead of just a fixed
background. String theory is a promising quantum gravity theory and in
this theory $p$-branes
emerge as a natural extension of point-like and string-like objects.
Black $p$-branes
\cite{horowitz:1991}, therefore, naturally extend black holes and thermodynamic properties
thereof to high dimensions. There are certain attempts to understand the origin of
Bekenstein-Hawking entropy of black branes with the knowledge gained in string theories
from a statistical point of view \cite{gubser:1996}.

Apart from the efforts on trying to find out the microscopic
explanation of black hole entropies, more thorough inspection on black
hole thermodynamic laws \cite{bardeen:1973} and their phase structures
has been proposed. A pioneering method that was put forward in
\cite{gibbons:1977}  is to evaluate the Euclideanized Einstein-Hilbert
action with the classical black hole solution as a zeroth order
approximation to the partition function and has been successfully
applied to specific gravitational solutions such as AdS black holes
\cite{hawking:1983}, Schwarzschild black
holes \cite{york:1986}, charged (AdS or RN) black holes
\cite{whiting:1988,Braden:1990hw,chamblin:1999,chamblin:1999-2,carlip:2003,lundgren:2008,banerjee:2011,banerjee:2012}
and even Kerr-Newman-AdS black holes \cite{caldarelli:1999} and
Gauss-Bonnet black holes \cite{cai:2002, cai:2007}.  Among these
efforts, asymptotically flat black holes, due to their negative
specific heat and Hawking radiation, may not have a well-defined
canonical ensemble description.
Hence, in \cite{york:1986,whiting:1988}, the authors proposed to
hypothetically put the Schwarzschild black hole in a cavity with a
constant temperature to form a canonical ensemble and then thermally
stable black holes with positive specific heat can also be found.
This analysis has also been extended to asymptotically flat black holes
with charges \cite{Braden:1990hw,carlip:2003}. For charged black
holes, there can be two kinds of boundary (i.e. the cavity)
conditions, either fixing the charge within the cavity, which
corresponds to the canonical ensemble, or fixing the electric
potential at the boundary, which corresponds to the grand canonical
ensemble. Depending on which ensemble we examine, the resulting phase
structure of the black hole can be quite different. In grand canonical
ensemble, since the charge is not fixed, there could be, in principle,
a black hole phase or a hot flat space phase. If the temperature and
electric potential at the boundary are carefully chosen, there could
be a first order phase transition between these two phases because
they have the same free energy (classical action). This first order
phase transition is the analogue of the Hawking-Page phase transition
in an AdS black hole system \cite{hawking:1983}. In a canonical
ensemble, since the charge is fixed and no space can be both flat and
charged, this Hawking-Page-like phase transition cannot happen in this
case. However, there may be a van der Waals-like phase transition
which occurs between two black hole phases of different size and ends
at a critical point where a second order phase transition takes place.

All the above analysis can also be applied to black brane/bubble \cite{witten:1982}
systems\footnote{We are here considering only the thermodynamic stability of
the brane system in a cavity. The dynamical stability, or the so-called
``Gregory Laflamme instability''\cite{Gregory:1993vy}, of the
chargeless black branes in a cavity 
was also studied in \cite{Emparan:2012be} and  was found to be
correlated  
with the thermodynamic stability. } which embody richer structure \cite{lu:2011,lu:2012-2,lu:2011-2,wu:2012}
since the spacetime dimension in superstring theory is ten and there could be branes
of $p+1$ dimensions where $p=0,1,2,\cdots$. Almost in all these studies, Hawking-page-like
or van der Waals-like phase transitions are found (except for some special $p$).
There can also be combinations of black branes with different dimensions, i.e.
\Dp-D$q$-branes ($q>p$) where \Dp-branes are uniformly smeared on D$q$-branes. In
our last paper \cite{zhou:2015}, we did a thorough scan over \Dp-\Dppf-brane systems ($p=0,1,2$)
on their thermal structures in various ensembles, which is a natural extension of
the work done by Lu et al. \cite{lu:2012}. In Lu's paper, the phase structure and
critical behavior of \Dp-\Dppf-brane system, especially the D1-D5 system, in canonical
ensemble is elaborately studied. Then a rather direct question is what their phase
structures are in other ensembles. Now that \Dp- and \Dppf-branes coexist, each of
them can have its own canonical ensemble and grand canonical ensemble. So there are
another three different ensembles:
\Dp\ in canonical ensemble
\Dppf\ in grand canonical ensemble (CG ensemble), \Dp\ in grand canonical ensemble \Dppf\ in
canonical ensemble (GC ensemble), and both \Dp\ and \Dppf\ in grand canonical ensemble
(GG ensemble). In GG ensemble, the Hawking-Page-like phase transition is found in
all D0-D4 D1-D5 and D2-D6 systems while no critical behavior would happen in these
systems, which is consistent with those studies on black holes or black $p$-branes.
In CG or GC ensemble, the van der Waals-like phase transition and critical behavior
can only happen in the D0-D4 system, whereas in CC ensemble this feature can appear
in both D0-D4 and D1-D5 systems but not in the D2-D6 system. We also noticed an interesting
symmetry of interchanging the roles between \Dp- and \Dppf-branes. More precisely,
the phase structure remains unchanged under the following simultaneous transformations,
\[ {\rm D}p\ {\rm charge} \leftrightarrow {\rm D}(p+4)\ {\rm charge},\qquad
{\rm D}p\ {\rm potential} \leftrightarrow {\rm D}(p+4)\ {\rm potential}. \]

Although all results obtained from studies on all kinds of black holes and branes
seem to match very well, we have to point out that almost all these analyses only
concern about thermal stability conditions for corresponding systems,
which requires the stable phases to 
minimize the free energy which is equivalent to the positive specific
heat condition.  It is true that for chargeless systems, when there is
only one independent thermodynamic variable, i.e. temperature or entropy, thermal
stability is the same thing as thermodynamic stability. However, for charged
system with a second independent thermodynamic variable, i.e. charge
or electric potential, the thermodynamic stability also involves electrical stability which
in general should not be omitted, although in some special cases it can be shown
that being thermally stable implies being electrically stable \cite{lu:2011-2}.
In general, the stability anlysis follows from the second law of
thermodynamics, which is  related to the second order variations of the
thermodynamic potentials. Studying the full thermodynamic stability of a system
with more than one independent variables involves computations on
the positivity property of the Hessian matrix of the thermodynamic
potential evaluated at the stationary point in
the parameter space, e.g. the
``temperature-potential'' space for grand canonical ensembles
\cite{Braden:1990hw,lu:2011-2}) which may be complicated. Nonetheless,
these conditions can be reduced to the positivity of some gereralized response
funtions as was done in \cite{chamblin:1999-2} where 
the stability conditions of the Einstein-Maxwell-anti-de-Sitter(EMadS)  black hole reduce to the positivity of
specific heat and  ``isothermal permitivity''. While it is well-known
that the
positivity of the specific heat of the black hole indicates the stability of the system
under the fluctuation of the horison size, the isothermal permitivity
is the stability of the system under the electrical fluctuation, such
as the potential or charge fluctuations.
In the present paper, we will mainly adopt the method in \cite{chamblin:1999-2} and
try to perform a systematic investigation on the more general thermodynamic
stability of \Dp-\Dppf\ systems by including the electrical stability. Since we have three independent
variables here, we would expect that there are three response
functions for each ensemble.  
As a result of our investigation, we find that electrical perturbations prohibit
the small thermally stable black branes to be electrically stable. That means,
the black brane with larger horizon is now the only thermodynamicaly stable phase,
so the van der Waals-like first order phase transition cannot occur any more,
and neither can the second order phase transition.

The paper is organized as follows. In section \ref{sect:setup}, we describe the basic setups
of the problem to solve and discuss the meaning of the thermal and
electrical stability 
in canonical and grand canonical ensemble. The main results of the thermal
stability discussion in \cite{zhou:2015}  are reviewed and the
formulae obtained previously and needed for the calculation in this
paper are also collected at the end of section  \ref{sect:setup}. We
then derive the thermodynamic stability criteria by incorporating the
electrical stability conditions and reduce them in section
\ref{sect:criteria}. In section \ref{sect:electrical} we find out the
electrical stability constraints on the parameters by using the
reduced criteria and then by combining them with the thermal stability
results, we accomplish the whole thermodynamic analysis in section
\ref{sect:thermo}. Finally,  section \ref{sect:conclude} is devoted to the
conclusion and discussion.

\section{The \Dp-\Dppf-brane system \label{sect:setup}}
\label{sec:background}

\subsection{The brane system}

A \Dp-\Dppf-brane system can be described by the following Euclideanized metric,
dilaton and form fields (see section 2 of \cite{zhou:2015} for more details),
\begin{eqnarray}
  ds^2 & = & \DM^{\frac{1-p}{4}} \DS^{\frac{p-7}{8}}
  \left( \DP dt^2 + \DM \sum^{p}_{i=1} dx_i^2
  + \DS \sum^{p+4}_{j=p+1} dx_j^2 \right) \nonumber\\
  & & + \DM^{\frac{p^2-1}{4(3-p)}} \DS^{\frac{p+1}{8}}
  \left( \frac{d\rho^2}{\DP\DM} + \rho^2 d\Omega_{4-p}^2 \right) ,\nonumber\\
  e^\phi &=& \DM^{\frac{p-1}{2}} \DS^{\frac{3-p}{4}} ,\nonumber\\
  A_{[p+1]} &=& -i \frac{\DP}{\DS} \left( \frac{\DS-\DM}{\DS-\DP} \right)^{1/2}
  dt \wedge dx_1 \wedge \cdots \wedge dx_p ,\nonumber\\
  A_{[p+5]} &=& -i \DP \left( \frac{1-\DM}{1-\DP} \right)^{1/2}
  dt \wedge dx_1 \wedge \cdots \wedge dx_{p+4} ,
  \label{eq:solution}
\end{eqnarray}
where
\begin{eqnarray}
  \DP &=& 1-\frac{x}{\tilde \rho} ,\qquad\quad \DM = 1-
\frac{q^2}{x\tilde \rho} ,\nonumber\\
  \DS &=& \frac{-2Q^2(1-1/\tilde \rho)+\DP+\DM+\frac 1{\tilde
\rho}\sqrt{(\bar \Delta_--\bar\Delta_+)^2+4Q^2\bar
\Delta_+\bar\Delta_-}}{2(1-Q^2)}\, ,\nonumber\\
\bar\Delta_+ &=&1-{x}\,,\quad \bar\Delta_- = 1-
\frac{q^2}{x}\,.
  \label{eq:deltas}
\end{eqnarray}
The functions defined in \eqref{eq:deltas} are expressed using the parameters of CC
ensemble such as $Q$ and $q$ which are reduced \Dp\ and \Dppf\ charge (densities).
For the grand canonical ensemble of \Dp-branes, we should use  $\Phi$ instead of
$Q$ while for the grand canonical ensemble of \Dppf-brane, we use $\varphi$ rather
than $q$. $\Phi$ and $\varphi$ are the corresponding conjugate
potentials for $Q$ and $q$ defined in \cite{zhou:2015}, which are
proportional to the form fields at the boundary. The relations between these parameters are summarized in section \ref{sec:collection}.
The other parameter $x\in (0,1)$ is the reduced size of the horizon, which is defined as
$x\equiv \rho_+^{3-p}/\rho_b^{3-p}$ where $\rho_+$ is the coordinate of  the
outer horizon and $\rho_b$ the coordinate of the boundary (cavity).
Similarly, $\tilde\rho\equiv(\rho/\rho_b)^{3-p}$, $Q$,
$q$, $\Phi$ and $\varphi$ are all rescaled to the same range $(0,1)$ to simplify
the analysis. The Euclideanized metric in \eqref{eq:solution} possesses another
property that the Euclidean time direction is periodic in order to
avoid the conical
singularity. The reduced time defined as $t/(4\pi\rho_b)$ has a period
seen at the boundary
\begin{eqnarray}
  b = \frac{x}{3-p} \left( \frac{\bar\Delta_+}{\bar\Delta_-} \right)^{1/2}
  \left( 1- \frac{\bar\Delta_+}{\bar\Delta_-} \right)^{\frac{p-2}{3-p}}
  \left( 1- \frac{\bar\Delta_+}{\bar\Delta_*} \right)^{1/2}
  \label{eq:temperature}
\end{eqnarray}
which turns out to be proportional to the inverse temperature of the black
branes measured at the boundary.

By using quantities in \eqref{eq:solution}, we can compute in each ensemble the
the thermodynamic potential which is a function of variables such as $x$,
$b$, $Q$ or $\Phi$ and $q$
or $ \varphi$, e.g. the free energy in canonical
ensemble in terms of $x$, $q$ and $Q$. 
Minimizing the thermodynamic potential as a univariate
function of $x$ by fixing the other parameters, which is equivalent to  only considering  the thermal
stability condition of the equilibrium, can provide us with information
about phase diagrams of the system which has already been presented in
\cite{zhou:2015}. 
However, we need to emphasize that we are now dealing with a system with three independent
thermodynamic variables, and in principle there should be three
stability condition, one thermal stability condition and two electrical
stability conditions for both charges.  
The phase diagram, unlike most planar phase diagrams we have seen in
textbooks, should also be 3-dimensional. Though we can draw 3D phase
diagrams on a 2D paper, it may look neater to present them in planar
diagrams with respect to two variables and to use algebraic inequalities
for the third one, which in our convention would always be the $b$
direction, to describe the region for the stable phases. Now we
summarize the basic results obtained in \cite{zhou:2015} below as a reference
for later analysis. However, we will only describe the main traits in those
results and will not be rigorous about specific values.

In GG ensemble, there exists a region in the $\varphi$-$\Phi$ plane,
where one of the black
brane and the hot flat space phase is the globally stable phase while the other is
locally stable. Which one is globally stable depends on the value of $b$, and
there is a specific value of $b$ such that these two phases have equal Gibbs free
energy, which indicates a Hawking-Page-like phase transition  happening at this
$b$. This first order phase transition could occur for all $p=0,1,2$.
In GC or CG ensemble, for $p=1,2$ the system either is unstable or has
a stable black brane 
phase, but neither the Hawking-Page-like phase transition nor the van der Waals-like
phase transition could happen in the phase diagram. However, for $p=0$ case it is
possible that in some region of the $Q$-$\varphi$ or $q$-$\Phi$ plane and at some
specific $b$ which depends on the value of the other two parameters, a van der
Waals-like first order phase transition would arise and as $(\varphi,Q)$ or $(\Phi,q)$
evolves this first order phase transition will eventually end up at a second order
phase transition point. In CC ensemble, the van der Waals-like phase transition cannot
happen only in the $p=2$ case. For both $p=0,1$, this liquid-gas-like phase transition
is found in certain region of the $q$-$Q$ plane  at some
specific $b$,
and terminates at a second order phase transition point. What we will
show later in this paper is that 
after the electrical stability is considered, the small
black brane phase in the van der Waals-like phase transition is not stable any more,
which actually rules out the possibility of this phase transition. Thus in that
case the only stable phase is the large black brane phase.

\subsection{Thermal stability and electrical stability}
To gain further understanding of the electrical stability, let us
review the physical interpretation of the thermal
stability/instability. We have mentioned the fact that black holes/branes have
negative specific heat and they radiate, which makes it impossible for them to be
self-perpetuating in asymptotically flat spacetimes. Due to this innate instability,
we have to stabilize the black system by placing it inside a homeothermal reservoir
which may compensate the thermal loss of the black system. So what we are dealing
with is such a system that the black brane keeps emitting energy to the outside and
sucking energy from the reservoir at the same time. When the amount it emits equals
the amount it sucks, the system can be possibly in equilibrium. However, the system can be
truly stable or meta-stable only when this equilibrium can be preserved under small fluctuations.
What our result (and most literature) reveals is that the existence of reservoir
does not necessarily guarantee the thermal stability of the system.
When the system has negative specific heat, it may be in an
instantaneous balance but will still collapse under arbitrarily
small fluctuations of the temperature or the energy and can not form
an equilibrium with the reservoir. In this case, a tiny fluctuation of the system state would
end up either with an explosion to the hot flat thermal gas (the horizon always emits
more than it swallows) or with the reservoir inevitably being engulfed by the black brane
horizon (the horizon always absorbs more than it ejects) if there is not a truly
stable state lying between these two fates. Nevertheless, by putting
the black holes/branes inside a reservoir, unlike putting them in the
infinite flat space, there really may exist some stable phases with positive specific heat.


For black hole system with charges, as Hawking radiation always exists, there is no reason to forbid the
horizon from emitting charged objects. Thus, we have to impose
another property on the cavity such that the idea of canonical ensemble makes sense.
That is, the cavity should constantly trade charged objects with the horizon in such
a way  that the total charge of the system is conserved when fluctuations
are not considered. In our case, since there exist both \Dp\ and \Dppf\ charges,
the cavity should be able to supply both charges.  
Given this imposition, now we can
discuss the stability under small fluctuations. In this context, the electrical
instability of the system means that however small the fluctuation of
the charges in the system is it will cause the
horizon to keep swallowing more or fewer charges than it admits from
the reservoir, and thus will be going farther away from the equilibrium.
Similarly, to realize a grand canonical ensemble for one of the brane
charges or both, we should assume that the reservoir has a mechanism to fix
the electrical potential at the boundary by exchanging charged objects
with the inside. 
Since the fluctuations of charges or potentials are always there, the
electrical stability condition must be considered in discussing the
phase structure of the black brane system in various ensembles.

\subsection{Collections of previous results}
\label{sec:collection}

Here we collect some results obtained in \cite{zhou:2015} for later reference.
\begin{itemize}
  \item \Dp\ charge:
    \begin{eqnarray}
      Q(x,\Phib,q) &=& \frac{\Phib(x^2-q^2)}{(1-\Phib^2)\sqrt{x(1-x)(x-q^2)}} ,\nonumber\\
      Q(x,\Phib,\vphib) &=& \frac{z\Phib}{(1-\Phib^2)\sqrt{1-z}} .
      \label{eq:Q-charge}
    \end{eqnarray}
    where $z\equiv z(x,\vphib)=x(1-\vphib^2)>0$.
  \item \Dppf\ charges:
    \begin{eqnarray}
      q(x,\vphib) &=& \frac{x\vphib}{\sqrt{1-z}} .
      \label{eq:q-charge}
    \end{eqnarray}
  \item \Dp\ potential:
    \begin{eqnarray}
      \Phi(x,Q,q) &=& \frac{\lambda+q^2-x^2}{2\sqrt{Q^2x(1-x)(x-q^2)}} ,\nonumber\\
      \Phi(x,Q,\vphib) &=& \frac{\eta-z}{2Q\sqrt{1-z}} .
      \label{eq:Phi-potential}
    \end{eqnarray}
    where 
  $\lambda\equiv\lambda(x,Q,q)=\sqrt{4Q^2x(1-x)(x-q^2)+(x^2-q^2)^2}$ and
  $\eta\equiv\eta(x,Q,\vphib)=\sqrt{z^2+4Q^2(1-z)}$ .
  \item \Dppf\ potential:
    \begin{eqnarray}
      \varphi(x,q) &=& q \sqrt{\frac{1-x}{x(x-q^2)}} .
      \label{eq:phi-potential}
    \end{eqnarray}
  \item Reciprocal of temperature $(4\pi\bar{\rho}_b)b=1/T\,$:
    \begin{eqnarray}
      \bCC \equiv b(x,Q,q) &=& \left(\frac{x^2-q^2}{x-q^2}\right)^{\frac{p-2}{3-p}}
      \frac{x\sqrt{(1-x)(x^2-q^2+\lambda)}}{\sqrt{2}(3-p)(x-q^2)} ,\nonumber\\
      \bCG \equiv b(x,Q,\vphib) &=& \frac{z^{\frac{p-2}{3-p}}}{\sqrt{2}(3-p)}
      \sqrt{x(1-z)\big(z+\eta\big)} ,\nonumber\\
      \bGC \equiv b(x,\Phib,q) &=& \frac{x}{(3-p)(x-q^2)} \left(\frac{x^2-q^2}{x-q^2}\right)^{\frac{p-2}{3-p}}
      \sqrt{\frac{(1-x)(x^2-q^2)}{1-\Phib^2}} ,\nonumber\\
      \bGG \equiv b(x,\Phib,\vphib) &=& \frac{xz^{\frac{p-2}{3-p}}}{3-p}
      \sqrt{\frac{(1-\vphib^2)(1-z)}{1-\Phib^2}} .
      \label{eq:b-temperature}
    \end{eqnarray}
  \item Entropy:
    \begin{eqnarray}
      S(x,Q,q) &=& \left(\frac{x^2-q^2}{x-q^2}\right)^{\frac{1}{3-p}}
      \sqrt{\frac{x(x^2-q^2+\lambda)}{2(x-q^2)}} ,\nonumber\\
      S(x,\Phib,q) &=& \left(\frac{x^2-q^2}{x-q^2}\right)^{\frac{1}{3-p}}
      \sqrt{\frac{x(x^2-q^2)}{(x-q^2)(1-\Phib^2)}} ,\nonumber\\
      S(x,Q,\vphib) &=& z^{\frac{1}{3-p}} \sqrt{\frac{x(z+\eta)}{2}} ,\nonumber\\
      S(x,\Phib,\vphib) &=& xz^{\frac{1}{3-p}} \sqrt{\frac{1-\vphib^2}{1-\Phib^2}} .
      \label{eq:S-entropy}
    \end{eqnarray}
\end{itemize}
The domain of each independent variable ($x$, $Q$, $q$, $\Phib$ or $\vphib$) has been
normalized to the unit interval $(0,1)$ except in some cases the upper bound of $x$
is restricted to a variable $x_{max}$($\leq 1$) which depends on $(\Phib,\vphib)$ 
or $(\Phib,q)$ in order to avoid  the naked singularity.

\section{Stability criteria for thermodynamic equilibrium\label{sect:criteria}}

In thermodynamics, the stability condition for an equilibrium is independent of the
ensemble, and can be  obtained  either by the maximization of the
entropy with fixed $E$, $Q$,  $q$, or by minimization of the energy
with fixed $S$, $Q$, $q$.  
For example, we use the minimization of the energy condition
\begin{eqnarray}
  \delta E - \bar T \delta S - \bar\Phi \delta Q - \bar\varphi \delta q &=& 0 ,\nonumber\\
  \delta^2 E - \bar T \delta^2 S - \bar\Phi \delta^2 Q - \bar\varphi \delta^2 q &>& 0 ,
  \label{eq:equi-condi}
\end{eqnarray}
where $E$ is the internal energy of the brane system, and $S$ is the
entropy. $\bar T$, $\bar \Phi$ and $\bar \varphi$ are the
corresponding temperature and potentials at the boundary fixed by the
reservoir, which play
the roles of the Lagrange multipliers. Using the first law of the
theromodynamics for the black brane $\delta E= T \delta S + \Phi
\delta Q + \varphi \delta q$, one can obtain the equillibrium
condition $T=\bar T$, $\Phi=\bar \Phi$, $\varphi=\bar \varphi$. 
Then the first equation above gives the second
order variation of internal energy,
\begin{eqnarray}
  \delta^2 E = \delta \bar T \delta S +\bar T \delta^2 S + \delta \bar\Phi \delta Q
    + \bar\Phi \delta^2 Q + \delta\bar \varphi \delta q + \bar\varphi \delta^2 q .
  \label{eq:E-2nd-var}
\end{eqnarray}
Inserting this into the inequality of \eqref{eq:equi-condi} gives the
stability condition at the equilibrium
\begin{eqnarray}
  \delta\bar T \delta S + \delta 
\bar\Phi \delta Q + \delta \bar\varphi \delta q > 0 .
  \label{eq:min-condi}
\end{eqnarray}
Since at the equilibrium, barred quantities equal the unbarred
quantities, one does not need to distinguish the barred and unbarred
quantities. 
However, in different context, we need to keep the bars to
distinguish the quantities fixed on the boundary and the quantities of
the black branes as  functions of the other variables. 
Equation (\ref{eq:min-condi}) will be the starting point of our
stability analyses in various ensembles.


In CC ensemble, we use $T$, $Q$ and $q$ as independent variables, then
\eqref{eq:min-condi} becomes
\begin{align}
  &\left(
  \begin{array}{ccc}
    \delta T & \delta q & \delta Q
  \end{array} \right) \left(
  \begin{array}{ccc}
    \big(\PP{S}{T}\big)_{Qq} & \big(\PP{S}{q}\big)_{TQ} & \big(\PP{S}{Q}\big)_{Tq} \\
    \big(\PP{\varphi}{T}\big)_{Qq} & \big(\PP{\varphi}{q}\big)_{TQ} &
\big(\PP{\varphi}{Q}\big)_{Tq} \\
    \big(\PP{\Phi}{T}\big)_{Qq} & \big(\PP{\Phi}{q}\big)_{TQ} & \big(\PP{\Phi}{Q}\big)_{Tq}
  \end{array} \right) \left(
  \begin{array}{c}
    \delta T \\ \delta q \\ \delta Q
  \end{array} \right)
> 0\, .
  \label{eq:CC-min-condi}
\end{align}
By using the Maxwell relations $\big(\PP{S}{q}\big)_{TQ}=- 
    \big(\PP{\varphi}{T}\big)_{Qq}$ and $
\big(\PP{S}{Q}\big)_{Tq}=-\big(\PP{\Phi}{T}\big)_{Qq}$, the positivity
of the above quadratic form is equivalent to 
 the positivity conditions
\begin{equation}
  \left( \PP{S}{T} \right)_{Qq} > 0 ,\ 
\left(\PP{\varphi}{q}\right)_{TQ}
> 0 ,\ \left|
  \begin{array}{cc}
     \big(\PP{\varphi}{q}\big)_{TQ} &
\big(\PP{\varphi}{Q}\big)_{TQ} \\
     \big(\PP{\Phi}{q}\big)_{TQ} & \big(\PP{\Phi}{Q}\big)_{Tq}
  \end{array} \right| > 0 .
  \label{eq:CC-positivity}
\end{equation}
These conditions can then be further reduced to
\begin{eqnarray}
  \left( \PP{S}{T} \right)_{Qq} > 0, \quad
\left(\PP{\Phi}{Q}\right)_{T\varphi} > 0
  ,\quad \left(\PP{\varphi}{q}\right)_{TQ} > 0 .
  \label{eq:CC-criteria}
\end{eqnarray}
Since the specific heat capacity is defined as $C_{Qq}=T\left( \PP{S}{T}
\right)_{Qq}$, the first condition is just the positivity of the
specific heat capacity. The other two conditions means the positivity
of the other two response functions of so-called ``permitivities'' for the two
kinds of the charges, similar to
the definitions in~\cite{chamblin:1999-2}.


By the same token, we can obtain criteria for equilibria in the other
ensembles. The main results are collected in table \ref{tab:criteria}.
To study the thermodynamic equilibrium criteria listed in Table
\ref{tab:criteria} directly is a little complicated. For example, there are
partial derivatives with $S$ fixed which is not easy to handle
directly. Since our expressions listed in section \ref{sec:collection}
explicitly depend on $x$, we will reduce these conditions to some
convenient form in favor of $x$ in order to use these equations in the
following subsections. 
\begin{table}[!t]
  \centering
  \begin{tabular}{c|c|c}
    \hline
    Ensemble & Independent variables & Criterion \\
    \hline
    CC & $T$, $Q$, $q$ & $ \big(\PP{S}{T}\big)_{Qq} > 0, \quad
    \big(\PP{\Phib}{Q}\big)_{T\vphib} > 0 ,\quad \big(\PP{\vphib}{q}\big)_{TQ} > 0 $ \\
    \hline
    GC & $T$, $\Phib$, $q$ & $ \big(\PP{S}{T}\big)_{\Phib q} > 0, \quad
    \big(\PP{Q}{\Phib}\big)_{Sq} > 0 ,\quad \big(\PP{\vphib}{q}\big)_{T\Phib} > 0 $ \\
    \hline
    CG & $T$, $Q$, $\vphib$ & $ \big(\PP{S}{T}\big)_{Q\vphib} > 0, \quad
    \big(\PP{\Phib}{Q}\big)_{T\vphib} > 0 ,\quad \big(\PP{q}{\vphib}\big)_{SQ} > 0 $ \\
    \hline
    GG & $T$, $\Phib$, $\vphib$ & $ \left(\PP{S}{T}\right)_{\Phib\vphib} > 0, \quad
    \big(\PP{Q}{\Phib}\big)_{Sq} > 0 ,\quad \big(\PP{q}{\vphib}\big)_{S\Phib} > 0 $ \\
    \hline
  \end{tabular}
  \caption{Criteria for equilibrium in various ensembles}
  \label{tab:criteria}
\end{table}

\subsection{Reduction for CC ensemble }

First, to make our computation under control, we prefer to use the variable $b$ instead of $T$.
Since we know that $T\sim 1/b$, the condition $\PP{S}{T}>0$ is the same as $\PP{S}{b}<0$. In CC
ensemble, we have
\begin{eqnarray}
  \left(\PP{S}{b}\right)_{Qq} = \PP{(S,Q,q)}{(b,Q,q)} = \PP{(S,Q,q)}{(x,Q,q)} \PP{(x,Q,q)}{(b,Q,q)}
  = \left(\PP{S}{x}\right)_{Qq} \Bigg/ \left(\PP{b}{x}\right)_{Qq} .
  \label{eq:CC-pS-pb}
\end{eqnarray}
The numerator in the above result is actually the product of $b(x,Q,q)$ and $f(x,Q,q)$, and the
latter is the function we have obtained in the last equation of
Eq.~(2.35) in our previous paper~\cite{zhou:2015}.
It can be shown that $f(x,Q,q)>0$ and obviously $b(x,Q,q)$ is positive, so we proved
\begin{eqnarray}
  \left(\PP{S}{b}\right)_{Qq} < 0 &\Leftrightarrow& \left(\PP{b}{x}\right)_{Qq} < 0 .
  \label{eq:CC-pb-px}
\end{eqnarray}
Here we can see that the positive specific heat condition is
equivalent to the stability condition we used in our previous paper.

Next we reduce the third condition in Table~\ref{tab:criteria}. We can rewrite the partial derivative
\begin{eqnarray}
  \left( \PP{\vphib}{q} \right)_{bQ} = \PP{(\vphib,b,Q)}{(\vphib,x,Q)}
  \PP{(\vphib,x,Q)}{(q,x,Q)} \PP{(q,x,Q)}{(q,b,Q)} = \left( \PP{b}{x} \right)_{Q\vphib}
  \left( \PP{\vphib}{q} \right)_x \Bigg/ \left( \PP{b}{x} \right)_{qQ} > 0 .
  \label{eq:CC-dphi-dq}
\end{eqnarray}
According to \eqref{eq:CC-pb-px}, the denominator is negative, and by using
\eqref{eq:phi-potential} we can easily prove  that the second factor in the numerator is positive; therefore for
\eqref{eq:CC-dphi-dq} to hold true, the other term in the numerator has to be negative,
\begin{eqnarray}
  \left( \PP{b}{x} \right)_{Q\vphib} < 0 .
  \label{eq:CC-3rd-condition}
\end{eqnarray}

Similarly, we can reduce the second condition in Table~\ref{tab:criteria} through the
same procedure,
\begin{eqnarray}
  \left( \PP{\Phib}{Q} \right)_{b\vphib} = \left( \PP{b}{x} \right)_{\Phib\vphib}
  \left( \PP{\Phib}{Q} \right)_{x\vphib} \Bigg/ \left( \PP{b}{x} \right)_{Q\vphib} > 0 .
  \label{eq:CC-dPhi-dQ}
\end{eqnarray}
Then one can show that the second term in the numerator is positive by
explicit computation, and at the same time the denominator is negative
due to \eqref{eq:CC-3rd-condition}. So, finally we get the last reduced stability condition,
\begin{eqnarray}
  \left( \PP{b}{x} \right)_{\Phib\vphib} < 0 .
  \label{eq:CC-2nd-condition}
\end{eqnarray}

\subsection{Reduction for GC ensemble}

First we reduce the second condition using the following identity,
\begin{eqnarray}
  \left( \PP{Q}{\Phib} \right)_{Sq} = \PP{(Q,S,q)}{(Q,x,q)} \PP{(Q,x,q)}{(\Phib,x,q)}
  \PP{(\Phib,x,q)}{(\Phib,S,q)} = \left( \PP{S}{x} \right)_{Qq}
  \left( \PP{Q}{\Phib} \right)_{xq} \Bigg/ \left( \PP{S}{x} \right)_{\Phib q} .
  \label{eq:GC-dQ-dPhib}
\end{eqnarray}
The first term in the numerator is positive as discussed in the previous subsection,
and the second term can be shown to be positive as well by explicitly using the first
expression in \eqref{eq:Q-charge}. Thus the following equivalent relations,
\begin{eqnarray}
  \left( \PP{Q}{\Phib} \right)_{Sq} > 0 \quad \Leftrightarrow \quad \left( \PP{S}{x} \right)_{\Phib q} > 0 .
  \label{eq:GC-2nd-condition}
\end{eqnarray}

Next we reduce the first condition  using the same trick,
\begin{eqnarray}
  \left( \PP{S}{b} \right)_{\Phib q} = \PP{(S,\Phib,q)}{(x,\Phib,q)} \PP{(x,\Phib,q)}{(b,\Phib,q)}
  = \left( \PP{S}{x} \right)_{\Phib q} \Bigg/ \left( \PP{b}{x} \right)_{\Phib q} .
  \label{eq:GC-dS-db}
\end{eqnarray}
Now if \eqref{eq:GC-2nd-condition} is already satisfied, then the above relation implies,
\begin{eqnarray}
  \left( \PP{S}{b} \right)_{\Phib q} < 0 \quad \Leftrightarrow \quad \left( \PP{b}{x} \right)_{\Phib q} < 0 .
  \label{eq:GC-1st-condition}
\end{eqnarray}
The right hand side of \eqref{eq:GC-1st-condition} is indeed the condition we examined
in our previous paper, which represents the thermal stability condition when the first
electric stability condition is satisfied.

The other electrical stability condition can be recast in the same
fashion,
\begin{eqnarray}
  \left( \PP{\varphi}{q} \right)_{b\Phib} = \left( \PP{b}{x} \right)_{\Phib\vphib}
  \left( \PP{\vphib}{q} \right)_x \Bigg/ \left( \PP{b}{x} \right)_{\Phib q} > 0 .
  \label{eq:GC-dphi-dq}
\end{eqnarray}
Now if we already solved the other two conditions, and since it can be readily seen that
$\PP{\vphib}{q}>0$, we would be left with the condition
\begin{eqnarray}
  \left( \PP{b}{x} \right)_{\Phib\vphib} < 0 .
  \label{eq:GC-3rd-condition}
\end{eqnarray}
We still need to bear in mind that, after we solved this inequality, we have to restate the result in terms of $x$,
$\Phib$ and $q$.

\subsection{Reduction for CG ensemble }

In CG ensemble, we first calculate $(\PP{q}{\vphib})_{SQ}$,
\begin{eqnarray}
  \left( \PP{q}{\vphib} \right)_{SQ} = \left( \PP{S}{x} \right)_{qQ}
  \left( \PP{q}{\vphib} \right)_x \Bigg/ \left( \PP{S}{x} \right)_{\vphib Q} > 0 .
  \label{eq:CG-dq-dphi}
\end{eqnarray}
Again the first term in the numerator is positive as argued before, and the second term
is also positive by \eqref{eq:q-charge}. So the denominator must be postive in order for
\eqref{eq:CG-dq-dphi} to hold, $\left( \PP{S}{x} \right)_{\vphib Q} >
0 $. We then calculate the thermal condition $(\PP{S}{b})_{Q\vphib}<0$,
\begin{eqnarray}
  \left( \PP{S}{b} \right)_{Q\vphib} = \left( \PP{S}{x} \right)_{Q\vphib}
  \Bigg/ \left( \PP{b}{x} \right)_{Q\vphib} < 0 .
  \label{eq:CG-dS-db}
\end{eqnarray}
From the condition analysed above, we see that the numerator is positive, so the denominator
has to be negative, $\left( \PP{b}{x} \right)_{Q\vphib} < 0$. Lastly, for the last condition
\begin{eqnarray}
  \left( \PP{\Phib}{Q} \right)_{b\vphib} = \left( \PP{b}{x} \right)_{\Phib\vphib}
  \left( \PP{\Phib}{Q} \right)_{x\vphib} \Bigg/ \left( \PP{b}{x} \right)_{Q\vphib} > 0
  \label{eq:CG-dPhi-dQ}
\end{eqnarray}
to hold, we need the first term in the numerator to be negative, i.e. 
$\left( \PP{b}{x} \right)_{\Phi\vphib}<0$, because the second term
is positive by \eqref{eq:Phi-potential} and the term in the denominator is negative by
\eqref{eq:CG-dS-db}.

\subsection{Reduction for GG ensemble}

In GG ensemble, for \Dp\ electrical stability condition to hold,
\begin{eqnarray}
  \left( \PP{Q}{\Phib} \right)_{Sq} = \left( \PP{S}{x} \right)_{Qq}
  \left( \PP{Q}{\Phib} \right)_{xq} \Bigg/ \left( \PP{S}{x} \right)_{\Phib q} > 0 ,
  \label{eq:GG-dQ-dPhi}
\end{eqnarray}
we need the denominator to be positive because both terms in the numerator are positive.
For the other electric stability condition to be true, i.e.
\begin{eqnarray}
  \left( \PP{q}{\vphib} \right)_{S\Phib} = \left( \PP{S}{x} \right)_{q\Phib}
  \left( \PP{q}{\vphib} \right)_x \Bigg/ \left( \PP{S}{x} \right)_{\Phib\vphib} > 0 ,
  \label{eq:GG-dq-dphi}
\end{eqnarray}
the denominator also has to be positive as a consequence of \eqref{eq:GG-dQ-dPhi}.
Then because the thermal condition can be rewritten as
\begin{eqnarray}
  \left( \PP{S}{b} \right)_{\Phib\vphib} = \left( \PP{S}{x} \right)_{\Phib\vphib}
  \Bigg/ \left( \PP{b}{x} \right)_{\Phib\vphib} < 0 ,
  \label{eq:GG-dS-db}
\end{eqnarray}
we find that the denominator needs to be negative by \eqref{eq:GG-dq-dphi}.

\subsection{Summary of the reduced stability conditions}
In summary, we list all the reduced stability conditions for these
ensembles below
\begin{eqnarray}
  \left(\PP{b}{x}\right)_{Qq} < 0 ,\,
  \left( \PP{b}{x} \right)_{Q\vphib} \Bigg|_{\vphib\to\varphi(x,q)} <
0 ,\,
  \left( \PP{b}{x} \right)_{\Phib\vphib}
\Bigg|_{\Phib\to\Phib(x,Q,q),\,\vphib\to\varphi(x,q)} < 0 \,,\,
\text{(CC)}\,;
\label{eq:cc-reduced}\\
  \left( \PP{b}{x} \right)_{\Phib q} < 0 ,\quad
  \left( \PP{b}{x} \right)_{\Phib\vphib} \Bigg|_{\vphib\to\varphi(x,q)} < 0 ,\quad
  \left( \PP{S}{x} \right)_{\Phib q} > 0 \,,\quad
\text{(GC)}\,;\label{eq:GC-reduced}
\\
  \left( \PP{b}{x} \right)_{Q\vphib} < 0 ,\quad
  \left( \PP{S}{x} \right)_{\vphib Q} > 0 ,\quad
  \left( \PP{b}{x} \right)_{\Phib\vphib}
\Bigg|_{\Phib\to\Phi(x,Q,\vphib)} < 0 \,,\quad 
\text{(CG)}\,;\label{eq:CG-reduced}
\\
  \left( \PP{b}{x} \right)_{\Phib\vphib} < 0 ,\quad
  \left( \PP{S}{x} \right)_{\Phib\vphib} > 0 ,\quad
  \left( \PP{S}{x} \right)_{\Phib q} \Bigg|_{q\to q(x,\vphib)} > 0 \quad
\text{(GG)}\,.\label{eq:GG-reduced}
\end{eqnarray}
In each set of the  conditions, the first one is the thermal stability
condition and the second and the third come from the electrical
stability conditions for D$(p+4)$  and D$p$ charges, respectively.
In fact, all these different sets of stability conditions are
equivalent since there are only three independent conditions and the
stability for an equilibrium state should be independent of ensemble.
From the deduction, these conditions all come from
\eqref{eq:min-condi}, and only we are
choosing different sets of convenient conditions for these ensembles.
We also see that some conditions are shared by different
ensembles and only are expressed in different independent variables.

\section{Electrical stability analyses\label{sect:electrical}}

In following sections, we will delve into the reduced criteria obtained in the last section
and find the more explicit electrical stability conditions in terms of variables such as
$x$, $\Phib$, $\vphib$, etc. Readers not interested in the deduction details can skip the
following four subsections and just jump to section~\ref{sec:elec-sum} where you can find
the final summarized results.

\subsection{GG ensemble\label{subsec:elect-GG}}

Now we turn to electrical stability analysis in GG ensemble. By using the last equation
of \eqref{eq:S-entropy}, it is easy to check the stability condition for \Dppf-brane
\begin{eqnarray}
  \left( \PP{S}{x} \right)_{\Phib\vphib} = \frac{4-p}{3-p}
  \sqrt{\frac{1-\vphib^2}{1-\Phib^2}} \, z^{\frac{1}{3-p}} > 0
  \label{eq:GG-dppf-condition}
\end{eqnarray}
is always true. The stability condition for \Dp-brane, albeit needing some tricky
rearrangements, can be proved to be true as well,
\begin{eqnarray}
  \left( \PP{S}{x} \right)_{\Phib q} = z^{\frac{1}{3-p}}
  \sqrt{\frac{1-\vphib^2}{1-\Phib^2}}
  \left[ \frac{(5-p)\vphib^2(1+\vphib^2)}{2(3-p)(1-x)(1-\vphib^2)}
  + \frac{2(4-p)+(5-p)\vphib^2}{2(3-p)} \right] > 0 .
  \label{eq:GG-dp-condition}
\end{eqnarray}
These two conditions are actually consistent with our intuition, i.e., as the radius
of horizon grows, the entropy (which is essentially the area of horizon) must grow. This
result also confirms the conclusion made in \cite{lu:2011-2}. That is, in GG ensemble
the thermodynamic stability is implied by sole thermal stability.

\subsection{GC ensemble}

Let us look at the stability condition for \Dppf-brane charges first,
\begin{eqnarray}
  \left( \PP{b}{x} \right)_{\Phib\vphib} = \frac{1}{2x(3-p)^2}
  \left( \frac{x^2-q^2}{x-q^2} \right)^{\frac{1}{3-p}}
  \frac{(3-p)q^2 + 2x - (5-p)x^2}{\sqrt{(1-x)(1-\Phib^2)(x^2-q^2)}} < 0 .
  \label{eq:GC-dppf-condition}
\end{eqnarray}
This implies
\begin{eqnarray}
  (5-p)x^2 -2x + (3-p)q^2 > 0 ,
  \label{eq:GC-dppf-imply}
\end{eqnarray}
which is solved by $x<x^-$ or $x>x^+$ where
\begin{eqnarray}
  x^{\pm} = \frac{1 \pm \sqrt{1+(5-p)(3-p)q^2}}{5-p} .
  \label{eq:GC-dppf-x}
\end{eqnarray}
Obviously $x^-<0$, so we only need to consider $x>x^+$ case. Also we have
another condition that \cite{zhou:2015}
\begin{eqnarray}
  0< q < x < x_{max} = \frac{1-\Phib^2 + \sqrt{(1-\Phib^2)^2 + 4q^2\Phib^2}}{2} < 1.
  \label{eq:GC-q-xmax}
\end{eqnarray}
Since we have
\begin{eqnarray}
  x^+ = \frac{1 + \sqrt{1-q^2 + (4-p)^2q^2}}{5-p} > \frac{q+\sqrt{(4-p)^2q^2}}{5-p} = q ,
  \label{eq:GC-x-lower-bound}
\end{eqnarray}
the lower bound of $x$ for \eqref{eq:GC-dppf-condition} to be true is $x^+$.
However, for $x^+<x<x_{max}$ to be a valid expression, we still need $x^+<x_{max}$.
By noticing that
\begin{eqnarray}
  x_{max} \Bigg|_{\Phib=\sqrt{\frac{3-p}{5-p}}} = x^+
  \label{eq:GC-xmax=x+}
\end{eqnarray}
and
\begin{eqnarray}
  \PP{x_{max}}{\Phib} = - \Phib \left( 1 +
  \frac{1-\Phib^2-2q^2}{\sqrt{(1-\Phib^2-2q^2)^2+4q^2(1-q^2)}} \right) < 0 ,
  \label{eq:GC-dxmax-dPhi}
\end{eqnarray}
we can conclude $x_{max}>x^+$ only when $\Phib<\sqrt{\frac{3-p}{5-p}}$.
So \eqref{eq:GC-dppf-condition} finally gives
\begin{eqnarray}
  x^+ < x < x_{max} \qquad \textrm{and} \qquad 0 < \Phib < \sqrt{\frac{3-p}{5-p}} .
  \label{eq:GC-dppf-final}
\end{eqnarray}

Next, we deal with the stability condition for \Dp-brane charges,
$\left( \PP{S}{x} \right)_{\Phib q}>0$. This inequality has already
been proved in \eqref{eq:GG-dppf-condition} where  \eqref{eq:q-charge}
is used to substitute for $q$.  Since in \eqref{eq:q-charge}, for arbitrary
$x\in (0,1)$ and  $\varphi\in (0,1)$, the range of $q$ is $0<q<x$,
this inequality is also satisfied in terms of $q$ within its domain
here.  
Hence we conclude this subsection with the statement that \eqref{eq:GC-dppf-final}
gives the final electrical stability condition in GC ensemble.

\subsection{CG ensemble}

In CG ensemble, let us solve the stability condition for \Dp-brane
charges first which is
\begin{eqnarray}
  \left( \PP{b}{x} \right)_{\Phib\vphib} = \frac{Qz^{\frac{p-2}{3-p}}}{(3-p)^2}
  \frac{2-(5-p)z}{\sqrt{2x(\eta-z)}} < 0 .
  \label{eq:CG-dp-condition}
\end{eqnarray}
This relation gives us
\begin{eqnarray}
  z > \frac{2}{5-p} \qquad \textrm{or} \qquad x > x_0 \equiv \frac{2}{(5-p)(1-\vphib^2)} .
  \label{eq:CG-dp-zx}
\end{eqnarray}
In fact, the condition \eqref{eq:CG-dp-condition} and
\eqref{eq:GC-dppf-condition} are the same condition expressed in
different variables, so the result \eqref{eq:CG-dp-zx}
 is consistent with 
\eqref{eq:GC-x-lower-bound}, which can be checked by using \eqref{eq:q-charge}. 
On the other hand, since $x<1$, the second relation above also indicates
\begin{eqnarray}
  1-\vphib^2 > \frac{2}{5-p} \qquad \textrm{or} \qquad \vphib < \sqrt{\frac{3-p}{5-p}} ,
  \label{eq:CG-dp-phi}
\end{eqnarray}
otherwise there is no such $x$ that \eqref{eq:CG-dp-condition} holds. Next we prove
that the stability condition for \Dppf-brane is again satisfied automatically, which
can be na\"ively argued by the same reason stated in subsection \ref{subsec:elect-GG}. First we
write this condition explicitly,
\begin{eqnarray}
  \left( \PP{S}{x} \right)_{Q\vphib} = 
  \frac{Q^2 z^{\frac{1}{3-p}}}{(3-p)\eta\sqrt{2x(z+\eta)}}
  \left[ 2(5-p)-(13-3p)z+\frac{(4-p)z(z+\eta)}{Q^2} \right] > 0 .
  \label{eq:CG-dppf-condition}
\end{eqnarray}
The validity of this inequality relies on the sign of the term inside
the square brackets
which we rewrite as a function $h$ of $z$ and $Q$,
\begin{eqnarray}
  h(z,Q) = 2(5-p)-(13-3p)z + (4-p)z \frac{z+\sqrt{z^2+4Q^2(1-z)}}{Q^2} .
  \label{eq:CG-h-zQ}
\end{eqnarray}
Then we have
\begin{eqnarray}
  \PP{h}{Q} = - \frac{2}{\eta Q^3} \big[ 2Q^2(1-z) + z^2 + z\eta \big] < 0 ,
  \label{eq:cg-dh-dQ}
\end{eqnarray}
which means
\begin{eqnarray}
  h(z,Q) > h(z,1) = (5-p)(2-z) > 0 .
  \label{eq:CG-h-sign}
\end{eqnarray}
This proves \eqref{eq:CG-dppf-condition} to be true.

\subsection{CC ensemble}

By using \eqref{eq:Phi-potential} \eqref{eq:phi-potential} and \eqref{eq:b-temperature},
we can write down the explicit expression of the stability condition
for \Dp-brane charges,
\begin{eqnarray}
  \left( \PP{b}{x} \right)_{\Phib\vphib} = \frac{Q}{(3-p)^2}
  \left( \frac{x^2-q^2}{x-q^2} \right)^{\frac{p-2}{3-p}}
  \frac{(3-p)q^2 + 2x - (5-p)x^2}{\sqrt{2x(x-q^2)(q^2-x^2+\lambda)}} <
0 \,,
  \label{eq:CC-dp-condition}
\end{eqnarray}
which is just \eqref{eq:GC-dppf-condition} or \eqref{eq:CG-dp-condition}  in terms of $Q$ and $q$. Similar to the GC ensemble, this condition would give us $x>x^+$ where $x^+$ is defined
in \eqref{eq:GC-dppf-x}. So the \Dp-brane stability condition restricts $x$ to the region
$x^+<x<1$.

Next we turn to the stability condition for \Dppf-brane charges. That is,
\begin{eqnarray}
  \left( \PP{b}{x} \right)_{Q\vphib} =
  \frac{h(x,Q,q) \sqrt{\lambda-x^2+q^2}}{4\sqrt{2}(3-p)^2 Qx(1-x)(x^2-q^2)\lambda\sqrt{x(x-q^2)}}
  \left( \frac{x^2-q^2}{x-q^2} \right)^{\frac{1}{3-p}} < 0 ,
  \label{eq:CC-dppf-condition}
\end{eqnarray}
where
\begin{eqnarray}
  h(x,Q,q) &=& (3-p)q^6 + q^4x (2 - 18Q^2 + 6pQ^2 - 11x + 3px + 18Q^2x - 6pQ^2x) \nonumber\\
  && +\, x^3 \big[ 2Q^2(1-x)(px-7x-2+2p) + 2x^2 - (5-p)x^3 \big] \nonumber\\
  && +\, q^2x^2 \big[ 2Q^2(1-x)(7x-px+11-5p) - 4x + (13-3p)x^2 \big] \nonumber\\
  && +\, (x^2-q^2) \big[ (3-p)q^2 + 2x - (5-p)x^2 \big] \lambda .
  \label{eq:CC-h-xQq}
\end{eqnarray}
Now that $x$, $q$ and the bounds of $x$ (i.e. $x^+$ and 1) are independent of $Q$,
if $\PP{h}{Q}<0$, we would have
\begin{eqnarray}
  h(x,Q,q) < h(x,0,q) = -2 \big[ (5-p)x^2 - 2x - (3-p)q^2 \big] (x^2-q^2)^2 ,
  \label{eq:CC-h-max}
\end{eqnarray}
which is negative due to \eqref{eq:CC-dp-condition}. This means \eqref{eq:CC-dppf-condition}
is true if $\PP{h}{Q}<0$. Using \eqref{eq:CC-h-xQq}, we have
\begin{eqnarray}
  \PP{h}{Q} &=& -4Q(1-x)x \Bigg( (x-q^2) \big[ (7-p)x^2 + 2(1-p)x - 3(3-p)q^2 \big] \nonumber\\
  && +\, \frac{(x-q^2)(x^2-q^2)\big( (5-p)x^2-2x-(3-p)q^2 \big)}{\lambda} \Bigg).
  \label{eq:CC-dh-dQ}
\end{eqnarray}
The second term in the large parentheses is, of course, positive, so next thing to do is to
prove that the first term is positive as well. Needless to say, we only need to focus on the term
in the square brackets, which we denote  as
\begin{eqnarray}
  h_1(x,q) \equiv (7-p)x^2 + 2(1-p)x - 3(3-p)q^2 .
  \label{eq:CC-h-1}
\end{eqnarray}
Then we have
\begin{eqnarray}
  \PP{h_1}{x} = 2(7-p)(x-\frac{p-1}{7-p}) > 2(7-p)(x^+ - \frac{p-1}{7-p}) .
  \label{eq:CC-dh1-dx}
\end{eqnarray}
For $p=0,1$, the above expression is obviously always positive, and for $p=2$,
\begin{eqnarray}
  \PP{h_1}{x} > 10 \left( \frac{1+\sqrt{1+3q^2}}{3} - \frac{1}{7} \right) > 110/21 .
  \label{eq:CC-dh1-dx-p2}
\end{eqnarray}
So we have $\PP{h_1}{x}>0$ and hence
\begin{eqnarray}
  h_1(x,q) > h(x^+,q) = 2(4-p)x^+(1-x^+) > 0\,.
  \label{eq:CC-h1>0}
\end{eqnarray}
Thus we have proved $\PP{h}{Q}<0$ and therefore \eqref{eq:CC-dppf-condition} always
holds true as long as condition \eqref{eq:CC-dp-condition} is true.

\subsection{Electrical stability summary}
\label{sec:elec-sum}

To summarize, we collect all the new constraints from the electrical
stability as follows,
\begin{itemize}
  \item GG ensemble: no more constraints.
  \item CG ensemble:
    \begin{eqnarray}
      x_0 < x < 1 \qquad \textrm{and} \qquad 0 < \vphib < \sqrt{\frac{3-p}{5-p}} ,
      \label{eq:CG-electrical}
    \end{eqnarray}
    where $x_0$ is defined in \eqref{eq:CG-dp-zx}.
  \item GC ensemble:
    \begin{eqnarray}
      x^+ < x < x_{max} \qquad \textrm{and} \qquad 0 < \Phib < \sqrt{\frac{3-p}{5-p}} ,
      \label{eq:GC-electrical}
    \end{eqnarray}
    where $x^+$ is defined in \eqref{eq:GC-dppf-x}.
  \item CC ensemble:
    \begin{eqnarray}
      x^+ < x < 1 .
      \label{eq:CC-electrical}
    \end{eqnarray}
\end{itemize}
Notice that all the three sets of new conditions in CG, GC and CC ensembles
are derived from the $\left(\frac {\partial b}{\partial x}\right)_{{
}_{\Phi\varphi}}$, which is just the thermal stability condition for the
GG ensemble. 

\section{Thermodynamic stability\label{sect:thermo}}

In this section, we will combine the results from electrical stability analysis
and the thermal stability results to find the full thermodynamic stability conditions
in each ensemble.

\subsection{GG ensemble}
\label{sec:thermo-GG}

Since in GG ensemble thermal stability always guarantees electrical stability, we
can infer its phase structure from the thermal stability conditions directly and
the results are listed in table \ref{tab:GG-phase-structure}.
\begin{table}[!t]
  \centering
  \begin{tabular}{c|c|c}
    \hline
    Globally & (Only) locally & \\
    stable phase & stable phase & \Raise{7pt}{Conditions} \\
    \hline
    \support Black brane & Hot flat space & $(\Phib,\vphib)\in\textrm{C}$ , $b_2<\bb<b_0$ \\
    \hline
    $^\dagger$ Black brane & & \\
    Hot flat space & \Raise{8pt}{N/A} & \Raise{8pt}{$(\Phib,\vphib)\in\textrm{C}$ , $\bb=b_0$} \\
    \hline
    \support & & $(\Phib,\vphib)\in\textrm{C}$, $b_0<\bb<b_1$ \\
    \cline{3-3}
    \support \Raise{13pt}{Hot flat space} & \Raise{13pt}{Black brane}
    & $(\Phib,\vphib)\in\textrm{B}$, $b_2<\bb<b_1$ \\
    \hline
    \support & & $(\Phib,\vphib)\in\textrm{A}$, $\bb>b_{unstable}$ \\
    \cline{3-3}
    \support Hot flat space & N/A & $(\Phib,\vphib)\in\textrm{B}\cup\textrm{C}$, $\bb>b_1$ \\
    \cline{3-3}
    \support & & $(\Phib,\vphib)\in\textrm{B}$, $b_{unstable}<\bb<b_2$ \\
    \hline
    \support & & $(\Phib,\vphib)\in\textrm{A}$, $\bb<b_{unstable}$ \\
    \cline{3-3}
    \support \Raise{11pt}{N/A} & \Raise{11pt}{Hot flat space} & $(\Phib,\vphib)\in\textrm{C}$, $\bb<b_2$ \\
    \hline
    \multicolumn{3}{l}{$^\dagger$\footnotesize There is a first order phase transition between
    black brane and hot flat space in this case.}
  \end{tabular}
  \caption{GG ensemble phase structure}
  \label{tab:GG-phase-structure}
\end{table}
\begin{figure}[!ht]
  \centering
  \includegraphics[width=.45\textwidth]{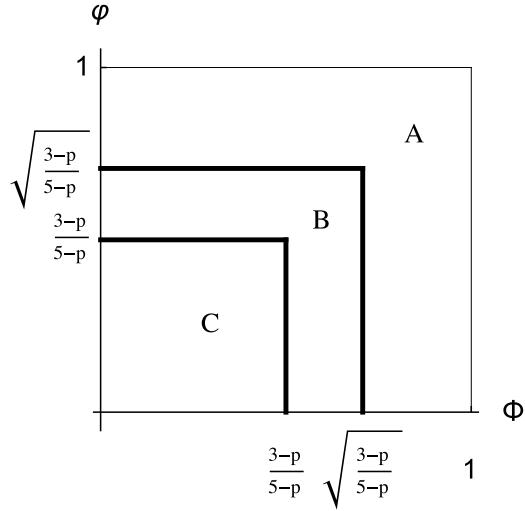}
  \caption{Regions in $\Phib$-$\vphib$ plane in GG ensemble.}
  \label{fig:GG-Phi-phi}
\end{figure}

The regions A, B and C  in this table are defined in figure~\ref{fig:GG-Phi-phi}
and other variables are defined as follows,
\begin{eqnarray}
  b_0 &=& \left[ \frac{4(4-p)}{(5-p)^2} \right]^{\frac{1}{3-p}} \frac{1}{(5-p)\sqrt{(1-\Phib^2)(1-\vphib^2)}} ,\nonumber\\
  b_1 &=& \frac{2^{\frac{1}{3-p}} (5-p)^{-\frac{5-p}{2(3-p)}}}{\sqrt{(3-p)(1-\Phib^2)(1-\vphib^2)}} ,\nonumber\\
  b_2 &=& \frac{\phi_{max}(1-\phi_{max}^2)^{\frac{1-p}{2(3-p)}}}{(3-p)\sqrt{1-\phi_{min}^2}} ,\nonumber\\
  b_{unstable} &=& \frac{(1-\phi_{max}^2)^{\frac{1}{3-p}}}{2(4-p)} \sqrt{\frac{1+\phi_{max}}{(1-\phi_{max})(1-\phi_{min}^2)}} ,
  \label{eq:GG-structure-variables}
\end{eqnarray}
where $\phi_{max/min}={\rm max/min}\{\Phib,\vphib\}$. Note that table~\ref{tab:GG-phase-structure}
is just tabularizing what we already obtained in \cite{zhou:2015} and we refer readers
to that paper for more details.

\subsection{GC ensemble}

\subsubsection*{D2-D6-branes}

\begin{figure}[ht]
  \centering
  \includegraphics[width=.4\textwidth]{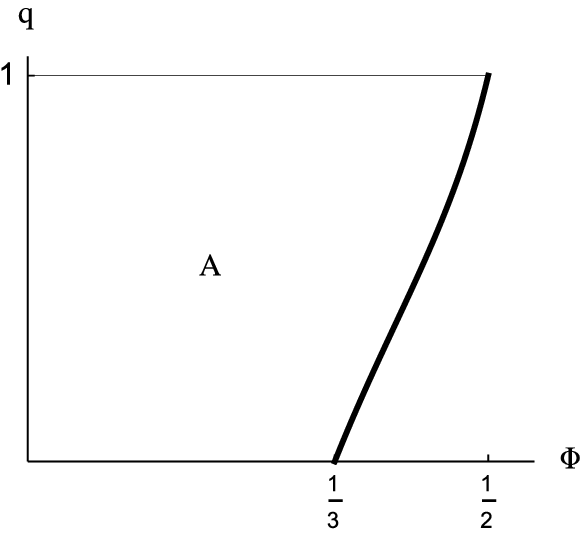}
  \includegraphics[width=.41\textwidth]{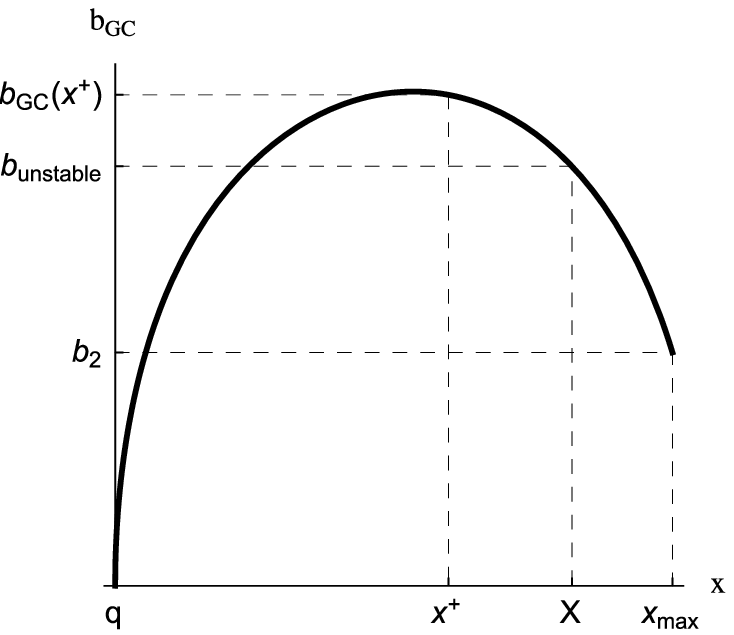}
  \caption{The first diagram shows the region in which black D2-D6 brane can be stable.
  The second shows the shape of $\bGC(x)$ and relations between $X$, $x^+$, $b_{unstable}$,
  $b_{GC}(x^+)$, etc.}
  \label{fig:GC-q-Phi-p2}
\end{figure}

According to our thermal stability analysis in our previous paper,
black branes can be thermally stable only when $(\Phib,q)$
lies in region A and $b_2<\bb<b_{unstable}$ where the right boundary of A is described
by the following equation,
\begin{eqnarray}
  q = \frac{2(1-\Phib)^2(3\Phib-1)}{\Phib(3-5\Phib)} ,\quad \frac{1}{3} < \Phib < \frac{1}{2} ,
  \label{eq:GC-Phib}
\end{eqnarray}
and $b_2=b_{GC}(x_{max})$. The other parameter $b_{unstable}$ is defined by $b_{unstable}=b_{GC}(X)$
where $X$ is the solution to equation\footnote{This equation is obtained by reducing (E.4)
in appendix E of \cite{zhou:2015} in $p=2$ case.} 
\begin{equation}
  9X^3 - (8-q+5q^2)X^2 + q^2(4+3q)X - 4q^3 = 0 .
  \label{eq:GC-X-p2}
\end{equation}
At $b_{unstable}$
the thermodynamic potential at the local minimum equals the one at
$x=q$. 
Region A in figure
\ref{fig:GC-q-Phi-p2}  automatically fulfils one of the electrical
stability conditions $\Phib<1/\sqrt{3}$.
The other electrical stability condition to be considered is
$x>x^+$. It can be proved  that at $x^+$, $db_{GC}/dx<0$ and $x^+$ lies on
the same decreasing branch of $b_{GC}(x)$ as $\bar x$ where the large
locally stable black brane lies.
One can then numerically show that $b_{GC}(x^+)>b_{unstable}$,
which implies $X>x^+$. Therefore, if $\bb$ lies between $b_2$ and $b_{unstable}$,
we will have $\xb>x^+$ (see the second diagram in figure \ref{fig:GC-q-Phi-p2}).
Thus we proved that the thermally stable black brane phase is also electrically
stable and is the only stable phase. The stability condition is
\begin{eqnarray}
  (\Phib, q) \in \textrm{A} \quad \textrm{and} \quad b_2<\bb<b_{unstable} .
  \label{eq:GC-p2-final}
\end{eqnarray}

\subsubsection*{D1-D5-branes}

\begin{figure}[!ht]
  \centering
  \includegraphics[width=.4\textwidth]{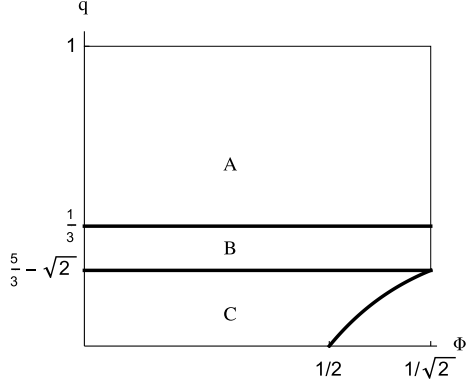}
  \caption{Regions in which there can be a stable D1-D5-brane system.}
  \label{fig:GC-q-Phi-p1}
\end{figure}

The electrical stability condition restricts $\Phib$ to
$(0,1/\sqrt{2})$ and thus provides a constraint to the results from the thermal stability
condition. With this constraint, in figure~\ref{fig:GC-q-Phi-p1}, the
region for black branes to be stable is
$(\Phib,q)\in\textrm{A}\cup\textrm{B}\cup\textrm{C}$, 
where the right boundary of C can be described by the following equation,
\begin{eqnarray}
  q = \frac{(2-\Phib)(2\Phib-1)}{3\Phib} ,\quad \frac{1}{2} < \Phib < \frac{1}{\sqrt{2}} .
  \label{eq:GC-q-Phi-p1}
\end{eqnarray}

\begin{figure}[!ht]
  \centering
  \includegraphics[width=.32\textwidth]{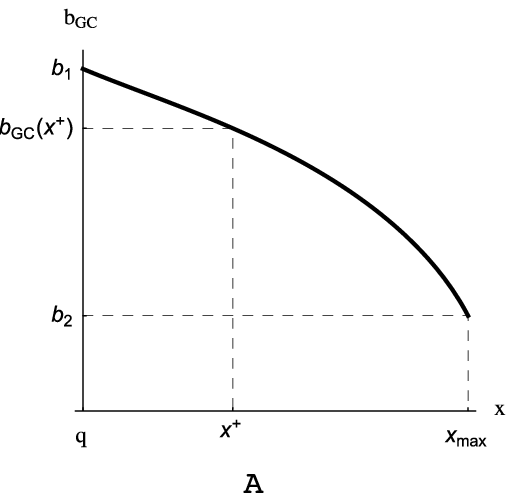}
  \includegraphics[width=.32\textwidth]{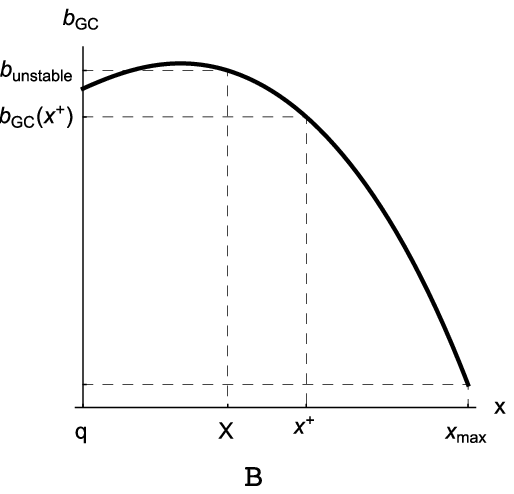}
  \includegraphics[width=.32\textwidth]{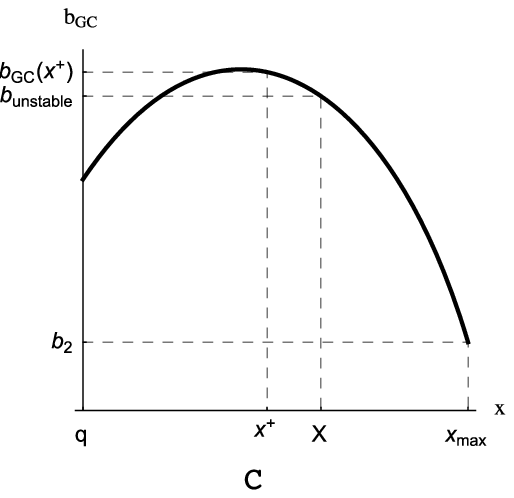}
  \caption{Typical shapes of $\bGC(x)$ in D1-D5 system. Diagrams labeled with A,B,C
  correspond to region A,B,C in figure~\ref{fig:GC-q-Phi-p1}, respectively.}
  \label{fig:GC-b-p1}
\end{figure}

In region A ($q>1/3$) of figure~\ref{fig:GC-q-Phi-p1}, $b_{GC}(x)$ is monotonically decreasing,
so the system has a thermally stable black brane if $b_2<\bb<b_1$ (see the first diagram
in figure~\ref{fig:GC-b-p1}). However, if $\bb>b_{GC}(x^+)$ (or $\xb<x^+$),
the system would be electrically unstable, hence the thermodynamic stability
requires $b_2<\bb<b_{GC}(x^+)$ in region A.

When $q<1/3$, the shape of $b_{GC}(x)$ looks like the one in the
second or the third diagram in
figure~\ref{fig:GC-b-p1} and there exists a $b_{unstable}$ above which the minimum
of the thermodynamic potential is at $x=q$ and there is no
globally stable black brane phase. In our previous
paper we have found that $b_{unstable}=b_{GC}(X)$ where\footnote{This solution is obtained
by solving (E.4) in appendix E of \cite{zhou:2015} in $p=2$ case.}
\begin{eqnarray}
  X = \frac{1}{8} \left( 3-2q+3q^2 + (1+q) \sqrt{3(3-q)(1-3q)} \right) .
  \label{eq:GC-X-p1}
\end{eqnarray}
When $b_2<\bb<b_{unstable}$, the black brane phase is thermally stable. Nonetheless,
we still have another condition for this system to be electrically stable, $\xb>x^+$.
Comparing $X$ with $x^+$, we find that when $q=q_0=\frac{5}{3}-\sqrt{2}$, we have
$X=x^+$, otherwise when $q>q_0$ (respectively, $q<q_0$), i.e. in region B (respectively,
region C) of figure~\ref{fig:GC-q-Phi-p1}, we have $X<x^+$ (respectively, $X>x^+$)
as shown in the second (respectively, the third) diagram in figure~\ref{fig:GC-b-p1}.

In summary, the thermodynamic stability condition for D1-D5 black
brane system is
\begin{eqnarray}
  (\Phib,q) \in \textrm{A} \cup \textrm{B} \cup \textrm{C} \quad \textrm{and} \quad
  b_2 < \bb < \left\{
    \begin{array}[]{ll}
      \bGC(x^+) , & (\Phib,q) \in \textrm{A} \cup \textrm{B} ,\\
      b_{unstable} , & (\Phib,q) \in \textrm{C}
    \end{array} \right. .
  \label{eq:GC-final-p1}
\end{eqnarray}

\subsubsection*{D0-D4-branes}

For D0-D4 system, the electrical stability condition requires
$0<\Phib<\sqrt{\frac{3}{5}}$ which further constraints the stability
regions resulted from the thermal stability analysis.
\begin{figure}[!ht]
  \centering
  \includegraphics[width=.4\textwidth]{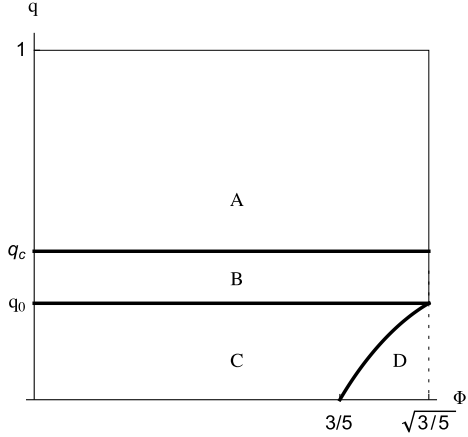}
  \caption{Regions in which there may exist a stable phase. The value of $q_c\approx 0.1416$
    was obtained in the previous paper and $q_0\approx 0.1227$ is obtained numerically.}
  \label{fig:GC-q-Phi-p0}
\end{figure}
As a result, in figure~\ref{fig:GC-q-Phi-p0}, the region of $0<\Phib<\sqrt{\frac{3}{5}}$ is then further assigned to four subregions, A, B, C and D, according to
the shapes of function $b_{GC}(x)$ 
in each subregion as shown in figure~\ref{fig:GC-b-p0}.
\begin{figure}[!h]
  \centering
  \includegraphics[width=.35\textwidth]{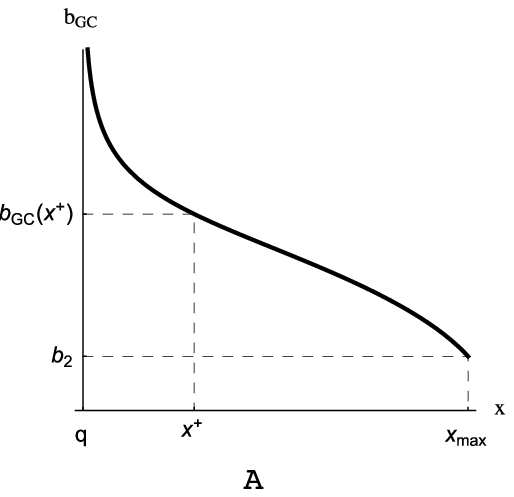} \quad
  \includegraphics[width=.35\textwidth]{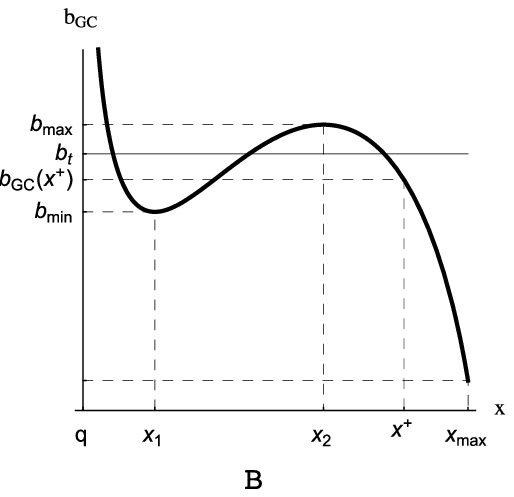} \quad
  \includegraphics[width=.35\textwidth]{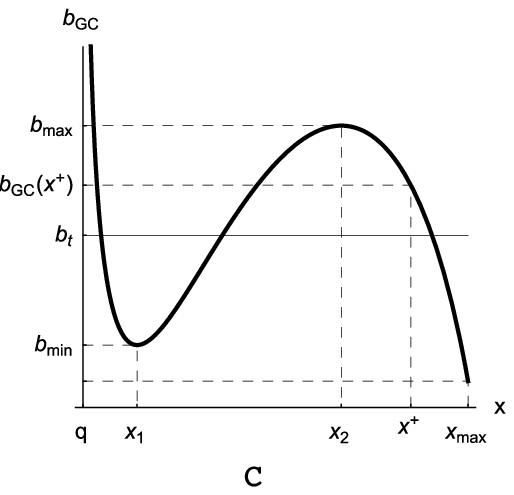} \quad
  \includegraphics[width=.35\textwidth]{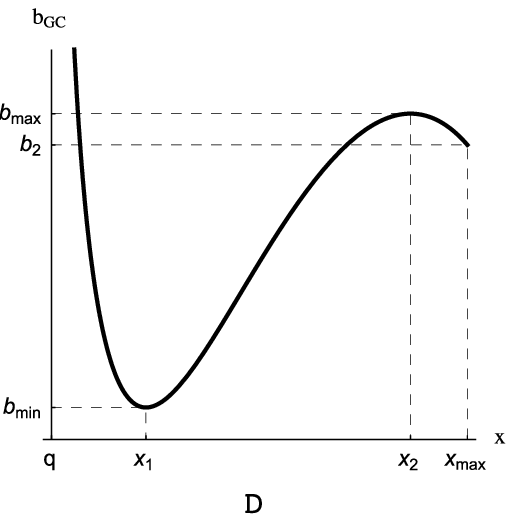}
  \caption{Four typical types of shapes of function $b_{GC}(x)$ in D0-D4 system. Diagram A/B/C/D
  correspond to region A/B/C/D in figure~\ref{fig:GC-q-Phi-p0} respectively.}
  \label{fig:GC-b-p0}
\end{figure}
The first diagram of figure~\ref{fig:GC-b-p0} corresponds to region A in the previous figure
where $b_{GC}$ is shown to be monotonically decreasing. So, at any $\bb>b_2$, the brane system
is thermally stable. In addition, electrical stability condition
requires $\xb>x^+$
or, equivalently, $\bb<b_{GC}(x^+)$. Therefore the condition for the brane phase to be stable in region A
is $b_2<\bb<b_{GC}(x^+)$.

\begin{figure}[!ht]
  \centering
  \includegraphics[width=.48\textwidth]{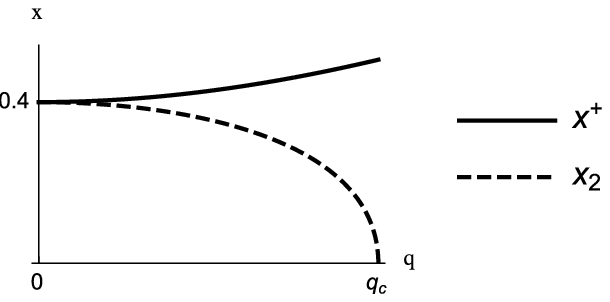}
  \includegraphics[width=.48\textwidth]{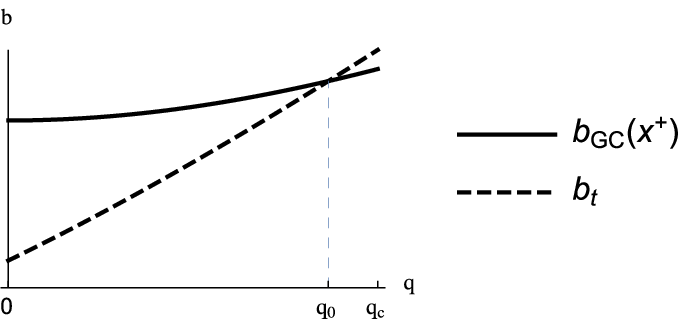}
  \caption{The first diagram shows the numerical result of $x^+>x_2$ in region B, C and D.
  The second diagram shows a general relation between $b_t$ and $\bGC(x^+)$ in region B and C
  though a special value $\Phib=0.5$ is chosen to plot this diagram.}
  \label{fig:GC-x+x<-p0}
\end{figure}

In region B, C or D, $b_{GC}$ has two extremum points, the minimum at
$x_1$  and the maximum at $x_2$.
At these two points, we define their corresponding values of $b_{GC}$ as $b_{min}=b_{GC}(x_1)$
and $b_{max}=b_{GC}(x_2)$. This shape of $\bGC$ makes it 
possible to have two locally stable black
brane phases, one with horizon size smaller than $x_1$ and the other
larger than $x_2$. However, the electrical stability requires $x>x^+$
and 
the numerical computation shows that $x^+>x_2$ as shown in the first
diagram in figure~\ref{fig:GC-x+x<-p0}, which means that the smaller black brane phase is by
no means electrically stable. So in the following analysis, we only need to find out whether
the larger black brane phases are stable. 

The difference between the $\bGC$ curves in region
B, C and D is that in region B or C there exists a temperature $T_t=1/b_t$ at which the large
black brane and the small one have the same thermodynamic potential whereas in region D, there is no such
$b_t$ because the large black brane, if exists, always has higher
thermodynamic potential than the smaller
one under the same temperature. So in region D the larger black brane is globally unstable which means there exist no
thermodynamicaly stable black brane phases in D, which is why we used
a dotted line as the right 
 boundary of region D in figure~\ref{fig:GC-q-Phi-p0}. The distinction between region
B and C will be dealt with as follows.
On the one hand, since in both regions, for $\bar b\in (b_2,b_t)$ ,
the larger black brane always has lower free energy than the
corresponding smaller black
brane (if there is one), the thermal stability condition would be $b_t>\bb>b_2$. On the other
hand, since $x^+>x_2$, the electrical stability condition now becomes $\bGC(x^+)>\bb>b_2$.
Combining these two conditions together, we get $b_2<\bb<\textrm{min}\{b_t,\bGC(x^+)\}$. So we need to
determine which one in $b_t$ and $\bGC(x^+)$ is smaller. Again by
numerical calculations, we
find that there exists a charge $q_0\approx 0.1227$, above which (in region B) we have
$\bGC(x^+)<b_t$ and below which (in region C) we have $\bGC(x^+)>b_t$,
which is illustrated in the second diagram in figure~\ref{fig:GC-x+x<-p0}. 
One may further notice that in figure~\ref{fig:GC-x+x<-p0}, near (but less than)
$q_c$ we have $\bGC(x^+)<b_t$, which means that the state  at the second order phase
transition point is electrically unstable. So, we have finally shown
that neither the second order
nor the first order phase transition could happen in GC ensemble.

To summarize, the final stability condition can be stated as follows:
the D0-D4 black brane can be a thermodynamically stable phase if
\begin{eqnarray}
  (\Phib,q) \in \textrm{A} \cup \textrm{B} \cup \textrm{C} ,\qquad
  b_2 < \bb < \left\{
    \begin{array}{ll}
      \bGC(x^+) , & (\Phib,q) \in \textrm{A} \cup \textrm{B}, \\
      b_t , & (\Phib,q) \in \textrm{C}
    \end{array} \right. ,
  \label{eq:GC-final-bb}
\end{eqnarray}
otherwise there is no known stable phase.

\subsection{CG ensemble}

The CG ensemble analysis is similar, though more complicated, to that in GC ensemble.
In fact, the symmetry of interchanging $(\Phib,q)$ and $(\vphib,Q)$ still exists after
the electrical stability conditions are considered.

\subsubsection*{D2-D6-branes}

\begin{figure}[!ht]
  \centering
  \includegraphics[width=.39\textwidth]{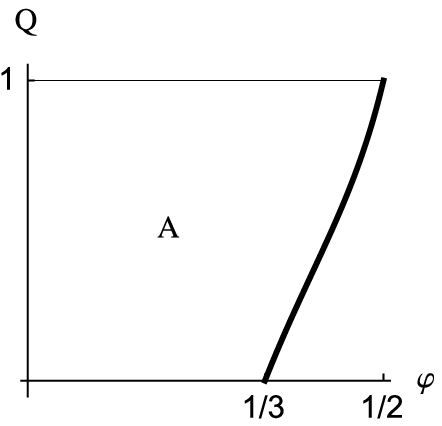} \quad
  \includegraphics[width=.4\textwidth]{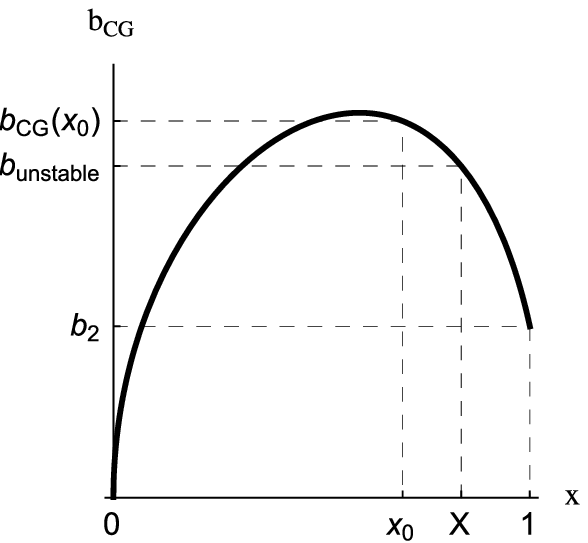}
  \caption{The first diagram shows the region in $Q$-$\vphib$ plane where the black brane
  can be stable and the second shows the shape of $\bCG$ in A.}
  \label{fig:CG-Q-phi-p2}
\end{figure}

By \eqref{eq:CG-electrical} and according to the thermal stability result, the constraint
on $(\vphib,Q)$ is shown in figure~\ref{fig:CG-Q-phi-p2} as region A whose right boundary is
\begin{eqnarray}
  Q = \frac{2(1-\vphib)^2(3\vphib-1)}{\vphib(3-5\vphib)} .
  \label{eq:CG-Q-phi-p2}
\end{eqnarray}
When $(\vphib,Q)\in\textrm{A}$, the shape of $\bCG(x)$ is shown in the second diagram of
figure~\ref{fig:CG-Q-phi-p2},
and there exists a $b_{unstable}=\bCG(X)$ below which (and above $b_2$) the black brane
phase has lowest thermodynamical potential. The critical value $X$ is defined as the solution to
the following equation (please refer to appendix E in \cite{zhou:2015}
for the detailed derivation),
\begin{eqnarray}
  f(X,Q,\vphib,p) \equiv \left( \frac{8-2p}{3-p} -Q \right) \sqrt{\xi} - \frac{(7-p)\xi}{2(3-p)}
  - \frac{3-\sqrt{(1-\xi)^2+4Q^2\xi}}{2} = 0 , \quad \textrm{for } p=2 ,
  \label{eq:CG-X}
\end{eqnarray}
where $\xi=1-X+X\vphib^2$. One can numerically show that
$x_0<X$ always holds, where $x_0$ is defined in \eqref{eq:CG-dp-zx}. So for any $\xb>X$, the black brane is automatically
electrically stable. Therefore, the thermodynamic stability condition is
\begin{eqnarray}
  (\vphib, Q) \in \textrm{A} \quad \textrm{and} \quad b_2<\bb<b_{unstable} .
  \label{eq:CG-p2-final}
\end{eqnarray}

\subsubsection*{D1-D5-branes}

\begin{figure}[!ht]
  \centering
  \includegraphics[width=.45\textwidth]{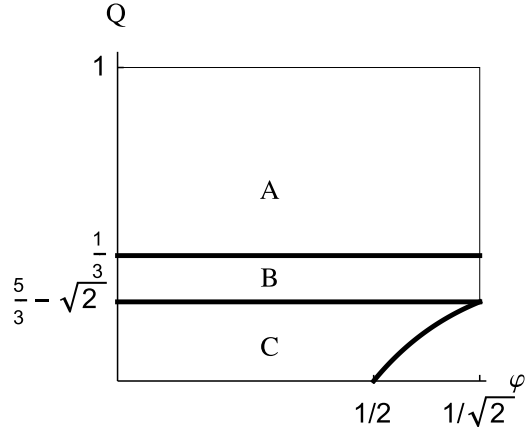}
  \caption{Regions in $Q$-$\vphib$ plane where there may exist a stable black brane phase.}
  \label{fig:CG-Q-phi-p1}
\end{figure}

\begin{figure}[!ht]
  \centering
  \includegraphics[width=.32\textwidth]{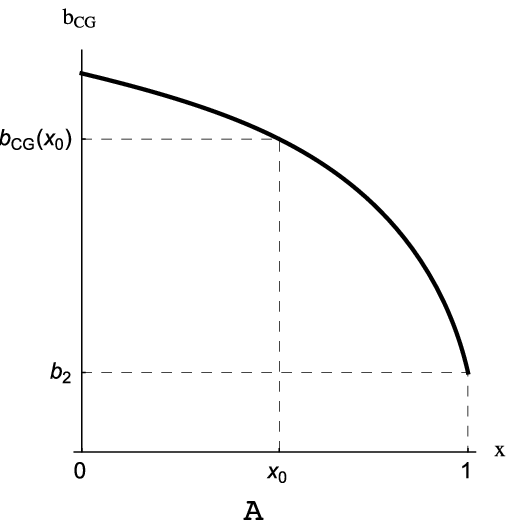}
  \includegraphics[width=.32\textwidth]{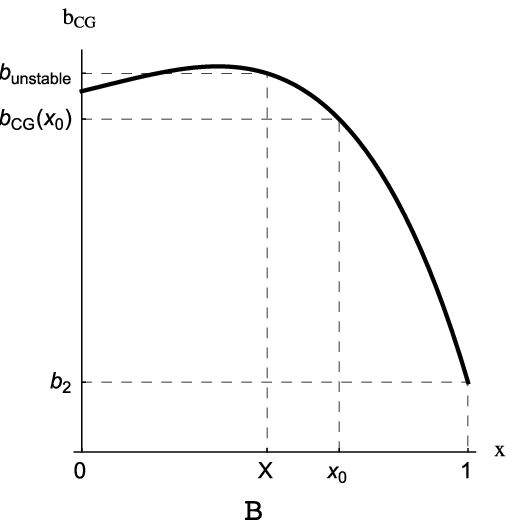}
  \includegraphics[width=.32\textwidth]{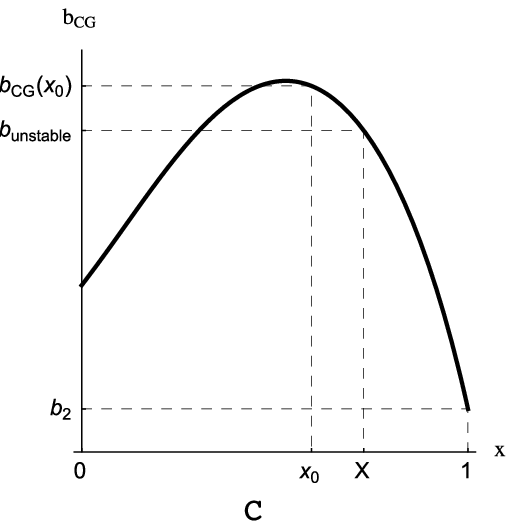}
  \caption{Typical shapes of $\bCG(x)$ in D1-D5 system. Diagram A/B/C correspond to region A/B/C in
  figure~\ref{fig:CG-Q-phi-p1} respectively.}
  \label{fig:CG-b-p1}
\end{figure}

Similar to GC ensemble, the electrical and thermal stability condition restricts $(\vphib,Q)$
to the region shown in figure~\ref{fig:CG-Q-phi-p1}.
The right boundary of region C in this figure is described by
\begin{eqnarray}
  Q = \frac{(2-\vphib)(2\vphib-1)}{3\vphib} .
  \label{eq:CG-Q-phi-p1}
\end{eqnarray}
In region A, $\bCG$ is monotonically decreasing and is depicted in the first diagram of
figure~\ref{fig:CG-b-p1}. It is easy to see that the stability condition for $\bb$ is
$b_2<\bb<\bCG(x_0)$. In region B or C, $\bCG$ could look like the curves in the second or
third diagrams depending on the values of $\vphib$ and $Q$ (actually only depending on
$Q$ which will be demonstrated later).
In both cases, there exists a $b_{unstable}$ and only below it (and above $b_2$) the black brane
can be thermally stable. Again $b_{unstable}=\bCG(X)$ where $X$ is the solution to $f(X,Q,\vphib,1)=0$.
In $p=1$ case, this equation can be reduced to the following equation,
\begin{eqnarray}
  2\xi - (5-3Q) \sqrt{\xi} + 2 = 0 ,
  \label{eq:CG-f-0-p1}
\end{eqnarray}
which has solution
\begin{eqnarray}
  \sqrt{\xi} = \frac{1}{4} \left( 5-3Q - \sqrt{3(3-Q)(1-3Q)} \right) .
  \label{eq:CG-f0-solution}
\end{eqnarray}
Now we want to find the region where $X<x_0$ which is equivalent to $\sqrt{\xi}>1/\sqrt{2}$.
This can be readily solved and gives $1/3>Q>5/3-\sqrt{2}$ which corresponds to region B in
figure~\ref{fig:CG-Q-phi-p1}. This also proves in region C we have $X>x_0$.

Combining all cases together, we obtain the final condition for D1-D5-branes to be stable
in CG ensemble,
\begin{eqnarray}
  (\vphib,Q) \in \textrm{A} \cup \textrm{B} \cup \textrm{C} \quad \textrm{and} \quad
  b_2 < \bb < \left\{
    \begin{array}[]{ll}
      \bCG(x_0) , & (\vphib,Q) \in \textrm{A} \cup \textrm{B}\, ,\\
      b_{unstable} , & (\vphib,Q) \in \textrm{C}\,.
    \end{array} \right. 
  \label{eq:CG-final-p1}
\end{eqnarray}

\subsubsection*{D0-D4-branes}

\begin{figure}[!ht]
  \centering
  \includegraphics[width=.4\textwidth]{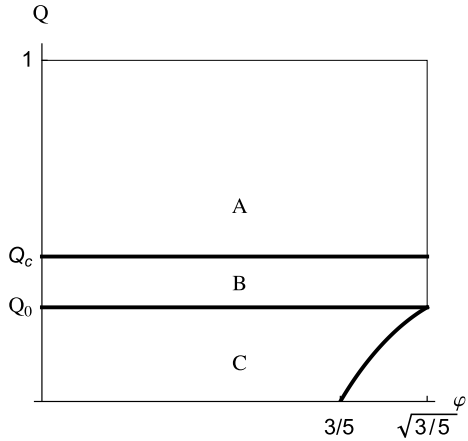}
  \caption{Regions in $Q$-$\vphib$ plane where stable phases can exist. $Q_c$ and $Q_0$
    respectively have the same values as $q_c$ and $q_0$ in GC ensemble, i.e.
    $Q_c\approx 0.1416$ and $Q_0\approx 0.1227$.}
  \label{fig:CG-p0-Q-phi}
\end{figure}

We have seen enough evidences  that GC ensemble and CG ensemble
are related to each other by the exchange of $\Phi\leftrightarrow\varphi$
and $Q\leftrightarrow q$, and  it is not too hard to convince oneself
that this is also true for D0-D4 branes by the same calculations as in
previous subsections. So, 
we shall omit most of the rigorous deduction details and simply
represent the final results below. The regions that can have both electrically and thermally stable
black brane phases in $Q-\vphib$ plane are shown in figure~\ref{fig:CG-p0-Q-phi}.
\begin{figure}[!ht]
  \centering
  \includegraphics[width=.32\textwidth]{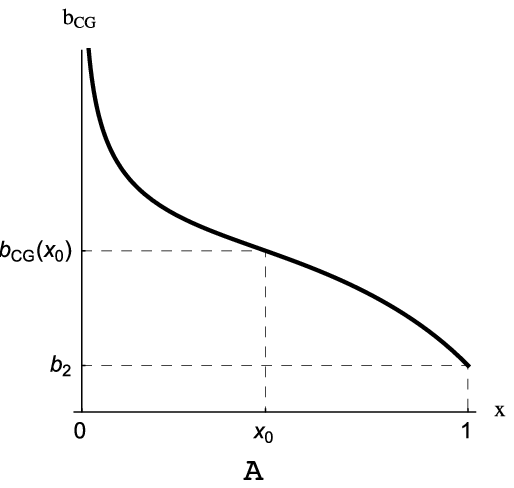}
  \includegraphics[width=.32\textwidth]{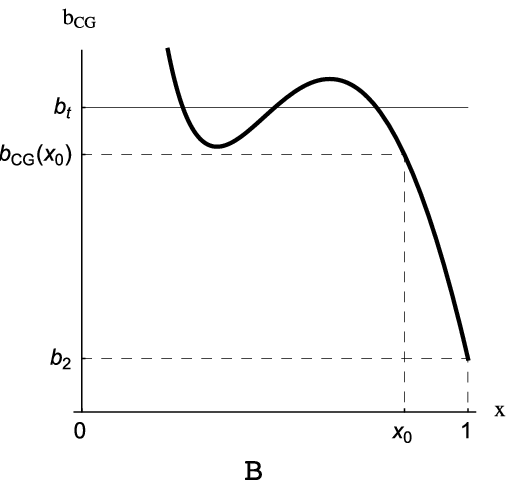}
  \includegraphics[width=.32\textwidth]{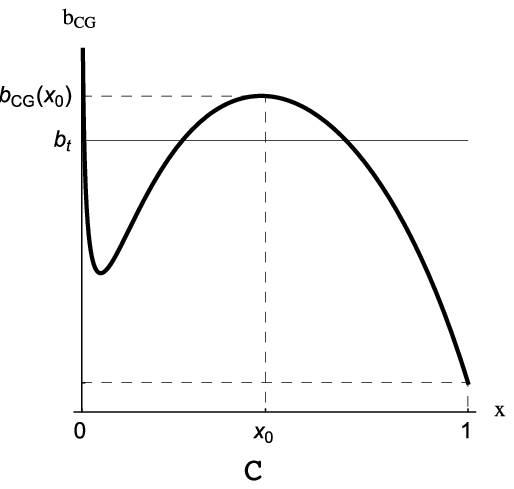}
  \caption{Typical shapes of $\bCG$ in D0-D4 system. Diagram A/B/C correspond to region A/B/C
  in figure~\ref{fig:CG-p0-Q-phi} respectively.}
  \label{fig:CG-p0-b}
\end{figure}
As to the constraint on $\bb$, in region A, $\bCG$ looks like the curve in the first
diagram in figure~\ref{fig:CG-p0-b}, so in order for the system to be stable, we need
$b_2<\bb<\bCG(x_0)$. In region B and C, $\bCG$ has the shape shown in the second and
third diagrams respectively and in both regions there exists a $b_t$ which has the
same meaning as the $b_t$ in figure~\ref{fig:GC-b-p0}. The difference between region
B and C is that in the former we have $b_t>\bCG(x_0)$ while in the latter otherwise.
Hence in region B, we need $b_2<\bb<\bCG(x_0)$ while in region C we need $b_2<\bb<b_t$.
Thus the final stability condition,
\begin{eqnarray}
  (\vphib,Q) \in \textrm{A} \cup \textrm{B} \cup \textrm{C} \quad \textrm{and} \quad
  b_2 < \bb < \left\{
    \begin{array}[]{ll}
      \bCG(x_0) , & (\vphib,Q) \in \textrm{A} \cup \textrm{B} ,\\
      b_t , & (\vphib,Q) \in \textrm{C}
    \end{array} \right. .
  \label{eq:CG-p0-final}
\end{eqnarray}
Again we would like to point out that the first/second order van der Waals-like phase
transition found by thermal stability computations can no long occur
now due to the
electrical instability.

\subsection{CC ensemble}

The thermal stability properties are gathered in the appendix A in our paper \cite{zhou:2015}.
These properties, especially in the D1-D5 case, are originally investigated in \cite{lu:2012}.
Now we can incorporate the thermal stability conditions there with the newly gained electrical
stability condition in \eqref{eq:CC-electrical} to find the thermodynamic stability conditions.

\subsubsection*{D2-D6-branes}

\begin{figure}[!ht]
  \centering
  \includegraphics[width=.5\textwidth]{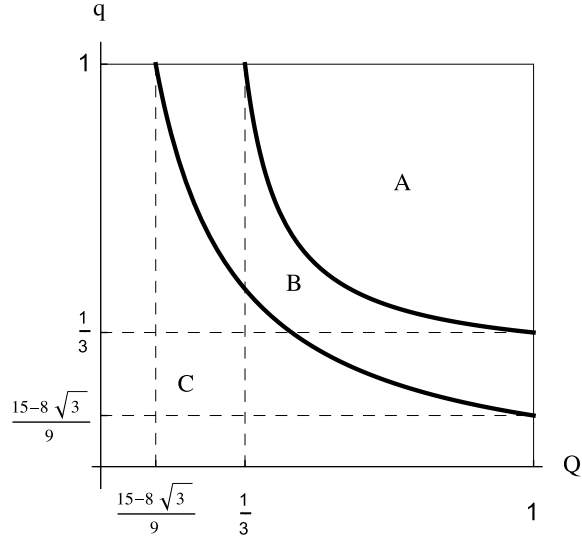}
  \caption{Regions in the $Q$-$q$ plane where there exist different types of $\bCC$ curves.}
  \label{fig:CC-p2-Q-q}
\end{figure}

\begin{figure}[!ht]
  \centering
  \includegraphics[width=.32\textwidth]{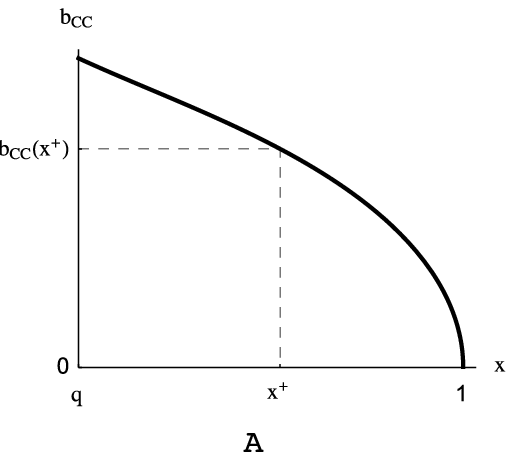}
  \includegraphics[width=.32\textwidth]{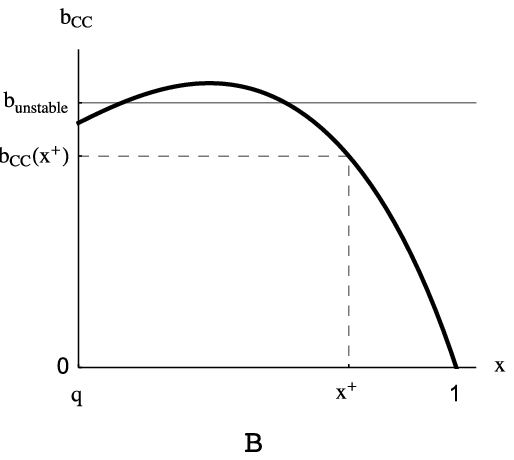}
  \includegraphics[width=.32\textwidth]{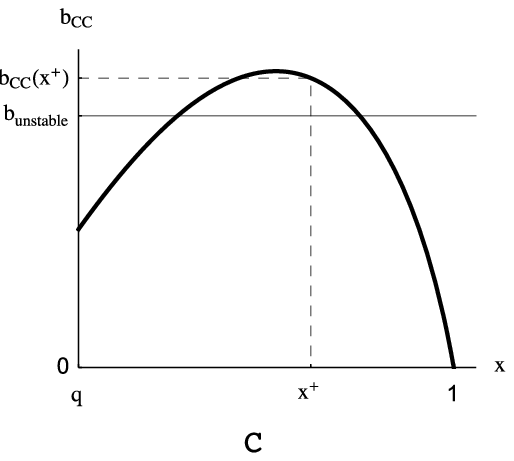}
  \caption{Shapes of $\bCC$ of D2-D6 system. Diagram A/B/C correspond to region
  A/B/C in figure~\ref{fig:CC-p2-Q-q} respectively.}
  \label{fig:CC-p2-b}
\end{figure}

Again we shall start with D2-D6 system in which almost all results can be obtained analytically.
In CC ensemble, the variable $b_2\equiv b(x_{max}=1)$ as defined in
other ensembles is just as 0. So
 for any high enough temperature (or small enough $\bb$) there always exists a stable (both
thermally and electrically) black brane phase. However, for large $\bb$ there can be very different
phenomena depending on the shapes of $\bCC$ in different regions of
$Q$-$q$ plane. In region A
(as shown in figure~\ref{fig:CC-p2-Q-q}), $\bCC$ is the curve shown in diagram A in figure~\ref{fig:CC-p2-b}.
One can see that in this region, the electrical stability constraint on $\bb$ is
$0<\bb<\bCC(x^+)$, where
\begin{eqnarray}
  \bCC(x^+) = \frac{1}{3\sqrt{3}} \sqrt{(1+\sqrt{1+3Q^2})(1+\sqrt{1+3q^2})} .
  \label{eq:CC-bplus}
\end{eqnarray}
The shapes of $\bCC$ in region B and C are shown in the other two diagrams B and C respectively.
In both B and C, there exists a $b_{unstable}$ that corresponds to the equality between the free energy
of black brane and the state with $x=q$, and only below this $b_{unstable}$ the black brane is the global
minimum of free energy. The difference between these two regions is that in region B, $b_{unstable}>\bCC(x^+)$
while in region C it is otherwise, and this difference obviously leads to different constraints
on $\bb$. The line which divides region B and C and also on which $b_{unstable}=\bCC(x^+)$ is described by
following relation,
\begin{eqnarray}
  (\sqrt{1+3q^2} - \sqrt{3}q) + (\sqrt{1+3Q^2} - \sqrt{3}Q) = 8 - 4\sqrt{3} .
  \label{eq:CC-BC-boundary}
\end{eqnarray}
There is also an interesting feature that the symmetry between D2 and D6 can be explicitly seen
via the symmetry between $Q$ and $q$ in \eqref{eq:CC-bplus} and \eqref{eq:CC-BC-boundary}.

To sum up, the dynamical stability condition for D2-D6 system in CC ensemble can be expressed
by the following simple constraint,
\begin{eqnarray}
  (0 <)\ \bb < \left\{
    \begin{array}[]{ll}
      \bCC(x^+) , & (Q,q) \in \textrm{A} \cup \textrm{B} ,\\
      b_{unstable} , & (Q,q) \in \textrm{C}
    \end{array} \right. .
  \label{eq:CC-p2-final}
\end{eqnarray}

\subsubsection*{D1-D5-branes}

\begin{figure}[!ht]
  \centering
  \includegraphics[width=.4\textwidth]{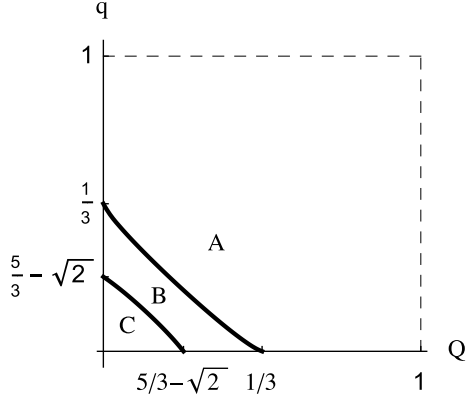}
  \caption{Regions in the $Q$-$q$ plane where there exist different types of $\bCC$ curves.}
  \label{fig:CC-p1-Q-q}
\end{figure}

\begin{figure}[!ht]
  \centering
  \includegraphics[width=.32\textwidth]{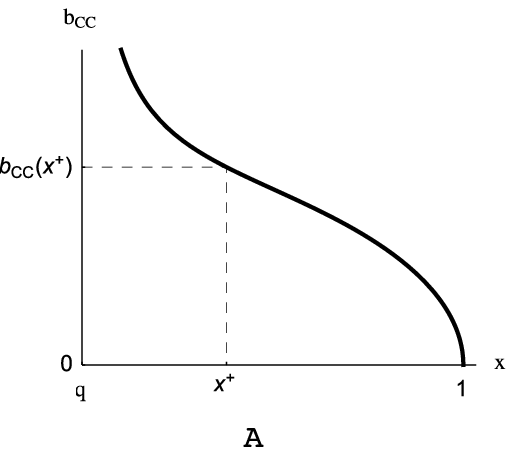}
  \includegraphics[width=.32\textwidth]{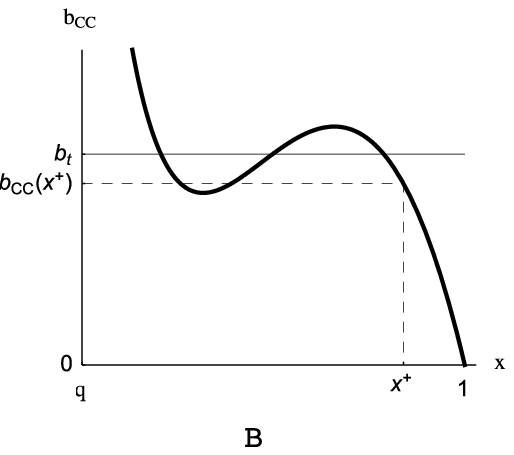}
  \includegraphics[width=.32\textwidth]{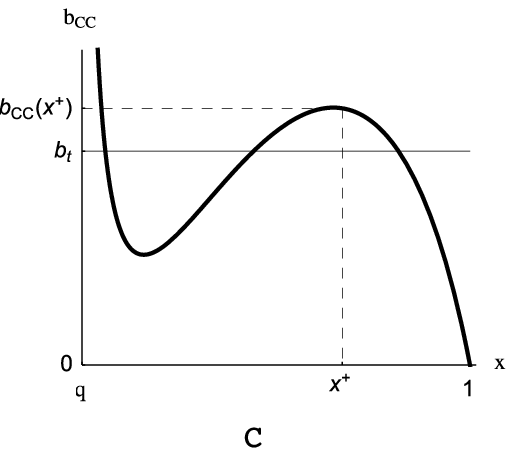}
  \caption{Shapes of $\bCC$ of D1-D5 (or D0-D4) system in CC ensemble. Diagram A/B/C
  correspond to region A/B/C in figure~\ref{fig:CC-p1-Q-q} respectively.}
  \label{fig:CC-p1-b}
\end{figure}

For D1-D5 system, the $Q-q$ plane can also be divided into three regions (see
figure~\ref{fig:CC-p1-Q-q}) and in each region the shape of $\bCC$ is shown
in figure~\ref{fig:CC-p1-b}.
The two lines dividing region A, B and B, C can not be analytically
solved and are both obtained numerically. 
With experiences of dealing with
so many cases in previous analyses, we can easily read out the final stability
condition from figure~\ref{fig:CC-p1-b},
\begin{eqnarray}
   (0 <)\ \bb < \left\{
    \begin{array}[]{ll}
      \bCC(x^+) , & (Q,q) \in \textrm{A} \cup \textrm{B} ,\\
      b_t , & (Q,q) \in \textrm{C}
    \end{array} \right. .
  \label{eq:CC-p1-final}
\end{eqnarray}

\subsubsection*{D0-D4-branes}

\begin{figure}[!ht]
  \centering
  \includegraphics[width=.4\textwidth]{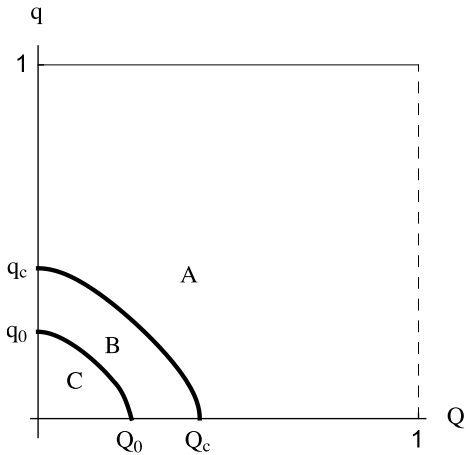}
  \caption{Regions in the $Q$-$q$ plane where there exist different types of $\bCC$
  curves. The constants are $q_c=Q_c\approx 0.1416$ and $q_0=Q_0\approx 0.1227$.}
  \label{fig:CC-p0-Q-q}
\end{figure}

The D0-D4 system has very similar feature to the D1-D5 system except that the division
of region A, B and C as shown in figure~\ref{fig:CC-p0-Q-q} is slightly different from
figure~\ref{fig:CC-p1-Q-q}.
The shapes of $\bCC$ in each region resemble those in figure~\ref{fig:CC-p1-b}. Therefore,
the final stability condition in terms of $\bb$ is exact the same as the condition in
\eqref{eq:CC-p1-final}, though the specific values of $\bCC(x^+)$ and $b_t$ are different
in general. One final remark is, in both D0-D4 and D1-D5 systems, the first order
and second order phase transitions would not arise any more because of the instability of
the small black brane phase in the first order phase transition and of
the state at the critical
point in the second order phase transition.

\subsection{Compatibilities}

Finally, we should check the compatibility of results obtained from different ensembles.
For example, if we take the limit $Q\to 0$ in CC ensemble, or we take the limit $\Phib\to 0$
in GC ensemble, in both limits we would end up with a canonical ensemble for \Dppf-branes.
Hence compatibility requires the stability conditions in these two ensembles should
degenerate into the same condition at the aforementioned limits. This can be easily
seen by noticing the following two facts. First, from \eqref{eq:b-temperature} we can
see that $b(x,Q=0,q)=b(x,\Phib=0,q)$ which means the thermal stability conditions coincides
in these two limits. Second, by setting $\Phib\to 0$, one realizes that in GC ensemble
$x_{max}\to 1$, so the two electrical stability conditions in \eqref{eq:GC-electrical}
and \eqref{eq:CC-electrical} degenerate. Therefore, we proved the thermodynamic stability
conditions in CC and GC ensembles are compatible. Similarly, one can also easily
check the other
compatibilities such as the degeneracy of CC and CG ensembles in the limit of $q\to 0$ and
$\vphib\to 0$, the degeneracy of GG and GC in the same limit and so on.  So, the conclusion is, the stability conditions in all ensembles are compatible with each other.

\section{Conclusions\label{sect:conclude}}

In this paper and our previous paper \cite{zhou:2015}, we have
conducted an exhaustive scan over the thermal and electrical
stabilities of \Dp-\Dppf-brane systems in all possible ensembles, and
this paper is especially focusing on the electrical stabilities.

We have confirmed in GG ensemble the thermal stability alone already
guarantees the thermodynamic stability which was found in
\cite{lu:2011-2} for \Dp-brane systems.  We find that in CG,GC and CC
ensembles, the electrical stability conditions generate extra
constraints on the horizon sizes and also on electric potentials only
in GC and  CG ensembles, besides the constraints that we already got
from thermal stability conditions. These new conditions rule out the
small brane phases from the old phase diagrams gained from pure
thermal stability considerations. Consequently, for all possible $p$,
there is no van der Waals-like phase structure any more, i.e. neither
the first order small-large black brane transitions nor the second
order phase transition. In the $Q\to0$ limit, the electrical stability
conditions also modify the one-charge brane system discussed in
\cite{lu:2011} such that the van der Waals phase structures are
invisible. Although we find this result in \Dp-\Dppf-brane systems, we
expect that it may be a common feature in other charged black hole
systems. In fact, this electrical instability has already been noticed
in EMadS black hole system studied in \cite{chamblin:1999-2} in which
the black holes with smaller horizons near the van der Waals phase
transition points were found to be unstable and hence the transition
does not exist after the electrical instability is considered. Whether
this is also the case for other systems could be left for future
research works.

From the new phase diagrams, we find that the symmetry of
interchanging $Q\leftrightarrow q$ and $\Phib\leftrightarrow\vphib$
is still kept after electrical stability conditions are considered. This
symmetry in fact could be a result of the T-duality\footnote{We would
like to thank 
Jun-Bao Wu for pointing out this possibility.}: If we make T-duality in
the $p+1,\dots,p+4$ directions of the D$(p+4)$ brane in which the
D$p$-charges are smeared, these two kinds
of branes interchange with each other, which is equivalent to make above
interchange in charges. This provides a natural interpretation of this symmetry
of the phase structure.

Now, we have included the thermal stability and the electrical
stability in the discussion of the thermodynamic structure of the black brane system.
However, there are another two independent parameters, the volume
of the cavity and the volume of the brane. The corresponding
generalized forces are the pressures on the cavity or in the brane
directions.  These parameters may also introduce new stability
conditions such as the compressibilities  in different directions.
This is in analogy to the similar discussions in AdS black holes or
asymptotically AdS black holes in which the cosmological constant
plays the role of the pressure \cite{Kastor:2009wy} and the positivity
of the compressibility also determines the stability of the
system\cite{Dolan:2014lea}. To discuss the effects of this kind of
instability on the phase structure of the black brane system could also
be a future research direction.

\section*{Acknowledgments}

This work is supported by the National Natural Science Foundation of
China under grant No.~11105138, and 11235010. Z.X. is also partly
supported by the Fundamental Research Funds for the Central
Universities under grant No.~WK2030040020. He also thanks Liang-Zhu Mu
and Jun-Bao Wu for helpful discussions. D.Z. is indebted to the
Chinese Scholarship Council (CSC). He would also like to thank Prof.
Yang-Hui He for providing him with a one-year visiting studentship at
City University London. Finally, the authors are specially grateful to
the referee of our last paper for the insightful suggestions to
initiate this work.

\end{document}